\documentclass{aa}

\usepackage{graphicx}
\usepackage{txfonts}
\usepackage{lscape}  %to rotate table
\usepackage{color}
\usepackage[colorlinks,citecolor=blue]{hyperref}

\usepackage{multirow}  % to combine rows and columns in table
\usepackage{xcolor}

\begin{document} 

\title{R-matrix electron-impact excitation data for the O-like iso-electronic sequence}

%\subtitle{TBA}

\author{Junjie Mao \inst{\ref{inst_strath}}
    \and N. R. Badnell \inst{\ref{inst_strath}}
    \and G. Del Zanna \inst{\ref{inst_damtp}}
    }

   \institute{Department of Physics, University of Strathclyde, Glasgow G4 0NG, UK\label{inst_strath}
   \and Department of Applied Mathematics and Theoretical Physics, University of Cambridge, Cambridge CB3 0WA, UK\label{inst_damtp}
   }

   %\date{Received September 15, 1996; accepted March 16, 1997}

% \abstract{}{}{}{}{} 
% 5 {} token are mandatory
 
\abstract
% context heading (optional), leave it empty if necessary  
{Astrophysical plasma codes are built on atomic databases. In the current atomic databases, $R$-matrix electron-impact excitation data of O-like ions are limited. The accuracy of plasma diagnostics with O-like ions depends on the availability and accuracy of the atomic data. This is particularly relevant in the context of future observatories equipped with the next generation of high-resolution spectrometers. }
{To obtain level-resolved effective collision strengths of O-like ions from \ion{Ne}{III} to \ion{Zn}{XXIII} (i.e., Ne$^{2+}$ to Zn$^{22+}$) over a wide range of temperatures. This includes transitions up to $nl=5d$ for each ion. We also aim to assess the accuracy of the new data, as well as their impact on solar atmosphere plasma diagnostics, compared to those available within the CHIANTI database.}
{A large-scale $R$-matrix intermediate coupling frame transformation calculations were performed systematically for the O-like iso-electronic sequence. For each ion, 630 fine-structure levels were included in both the configuration interaction target and close-coupling collision expansions.}
{The present results (energy levels, oscillator strengths, and effective collision strengths) of selected ions across the iso-electronic sequence are compared with those in archival databases and the literature. For selected ions across the iso-electronic sequence. We find general agreement with the few previous $R$-matrix calculations of collision strengths. We illustrate the improvements for a few solar plasma diagnostics over existing CHIANTI atomic models based on distorted wave data. The electron-impact excitation data are archived according to the Atomic Data and Analysis Structure (ADAS) data class {\it adf04} and will be available in OPEN-ADAS.}
{}
\titlerunning{Electron-impact excitation data for O-like ions}
\authorrunning{J. Mao et al.}       
\maketitle

\section{Introduction}
\label{sct:intro}
Spectral lines of O-like ions can be used to constrain the physical properties (e.g., temperature, density, elemental abundance) of astrophysical plasmas. For instance, in the active region of the solar corona, the \ion{Fe}{XIX} $\lambda1118$ emission line is observed to study the hot ($10^6$~K) plasma emission \citep{fel73,wan06,dza21b}. The \ion{Fe}{XIX} $13.42$~\AA, $13.52$~\AA, and $13.74$~\AA\ emission lines from the ground and metastable levels are also observed in the Sun \citep{phi82}. In the hot corona of the spectroscopic binary Procyon, the observed ground and metastable emission lines of \ion{S}{IX} $47.25$~\AA, $55.54$~\AA, and $56.33$~\AA\ were used for density diagnostics \citep{li13}. In the high-mass X-ray binary Vera X-1, \ion{Mg}{V} 9.81~\AA\ was detected in absorption, while \ion{Si}{VII} 7.06~\AA\ and \ion{S}{IX} 5.32~\AA\ was detected in emission among other lines of Ne, Mg, and Si \citep{gri17,ama21}. These lines reveal a highly variable, structured accretion flow close to the compact object \citep{gri17}, as well as a multi-phase wind of the
companion star \citep{ama21}. 

From a theoretical perspective, \citet{raj78} studied the density dependence of solar emission lines of O-like ions (\ion{Ne}{III}, \ion{Mg}{V}, \ion{Si}{VII}, \ion{S}{IX}, \ion{Ar}{XI}). \citet{mao17} presented density diagnostics with the ground and metastable absorption lines of Be-like to C-like ions in the context of ionized winds driven away from Active Galactic Nuclei. As key diagnostic transitions (in emission and absorption) are in general weak \citep{dza18a,mao17}, future observatories equipped with the next generation of high-resolution spectrometers \citep{smi16,bar18,cui20} are certainly required to observe these lines. 

Astrophysical plasma codes built on extensive atomic databases (e.g., ADAS\footnote{http://www.adas.ac.uk}, AtomDB\footnote{http://www.atomdb.org/Webguide/webguide.php}, CHIANTI\footnote{https://www.chiantidatabase.org/}, SPEX\footnote{https://www.sron.nl/astrophysics-spex}) enables plasma diagnostics with high-quality spectra obtained with current and future generations of high-resolution spectrometers. Continuous development of the atomic databases is essential because the current databases are neither complete nor accurate as we would wish. 

Electron-impact excitation is one of the fundamental atomic processes to determine the level population of an ion. Systematic $R$-matrix intermediate coupling frame transformation (ICFT) calculations of the electron-impact excitation data have been performed for many iso-electronic sequences (Li-, Be-, B-, F-, Ne-, Na-, and Mg-like) since 2007 \citep[see][for a review]{bad16}. Data for C-like and N-like iso-electronic sequences are recently presented in \citet{mao20c} and \citet{mao20n}. The present work focuses on the O-like iso-electronic sequence. $R$-matrix electron-impact excitation data are available only for a few O-like ions. The number of energy levels (thus the transitions among the levels) and temperature range of the effective collision strength vary among existing calculations. 

Here we present systematic $R$-matrix calculations for O-like ions from \ion{Ne}{III} to \ion{Zn}{XXIII} (i.e. Ne$^{2+}$ to Zn$^{22+}$). We obtain effective collision strengths among 630 levels over a temperature range spanning five orders of magnitude for each ion. We describe the $R$-matrix calculation in Section~\ref{sct:mo}. Results are summarized in Section~\ref{sct:res}. We compare the new data with results of previous works in \ref{sct:dis}. We also show the impact of new data on solar atmosphere plasma diagnostics. A summary is provided in Section~\ref{sct:sum}. Additionally, a supplementary package is available at Zenodo \citep{mao21z}, which includes the input files of the $R$-matrix calculations, atomic data from the present work, archival databases, and literature, as well as scripts used to create the figures presented in this paper. 

\section{Method}
\label{sct:mo}
We used the same method for the structure and collision calculations as described in \citet{mao20c} and \citet{mao20n} for C- and N-like ions. The main difference is that, for O-like ions, we included a total of 630 fine-structure levels in both the configuration-interaction target expansion and the close-coupling collision expansion. These levels arise from the 27 configurations listed in Table~\ref{tbl:cfg}. 

%-----------------------------Table Start----------------------------
\begin{table*}
\caption{List of configurations used for the structure and collision calculations. }
\label{tbl:cfg}
\centering
\begin{tabular}{cl|cl|cl}
\hline\hline
\noalign{\smallskip} 
Index & Conf. & Index & Conf. & Index & Conf. \\
\noalign{\smallskip} 
\hline
\noalign{\smallskip} 
1 & $2s^22p^4$ & 2 & $2s2p^5$ & 3 & $2p^6$ \\
\noalign{\smallskip} 
4 & $2s^22p^33s$ & 5 & $2s^22p^33p$ & 6 & $2s^22p^33d$ \\
\noalign{\smallskip} 
7 & $2s2p^43s$ & 8 & $2s2p^43p$ & 9 & $2s2p^43d$ \\
\noalign{\smallskip} 
10 & $2p^53s$ & 11 & $2p^53p$ & 12 & $2p^53d$ \\
\noalign{\smallskip} 
13 & $2s^22p^34s$ & 14 & $2s^22p^34p$ & 15 & $2s^22p^34d$ \\
\noalign{\smallskip} 
16 & $2s^22p^34f$ & 17 & $2s2p^44s$ & 18 & $2s2p^44p$ \\
\noalign{\smallskip} 
19 & $2s2p^44d$ & 20 & $2s2p^44f$ & 21 & $2p^54s$ \\
\noalign{\smallskip} 
22 & $2p^54p$ & 23 & $2p^54d$ & 24 & $2p^54f$ \\
\noalign{\smallskip} 
25 & $2s^22p^35s$ & 26 & $2s^22p^35p$ & 27 & $2s^22p^35d$  \\
\noalign{\smallskip} 
\hline
\end{tabular}
\end{table*}
%----------------------------Table End------------------------------

\subsection{Structure}
\label{sct:str}
The AUTOSTRUCTURE code \citep{bad11} is used for the target atomic structure calculation. By diagonalizing the Breit-Pauli Hamiltonian \citep{eis74}, wave functions are calculated. The one-body relativistic terms (mass-velocity, nuclear plus Blume \& Watson spin-orbit and Darwin) are included perturbatively. The Thomas-Fermi-Dirac-Amaldi model is used for the electronic potential with $nl$-dependent scaling parameters \citep{nus78} as shown in Table~\ref{tbl:sca_par}. These scaling parameters are obtained in the same way for all the ions along the iso-electronic sequence without further manual adjustment. 

As recognized in \citet{mao20c} and \citet{mao20n}, our calculation leads to a relatively poor structure for low-charge ions (e.g., \ion{Ne}{III} and \ion{Mg}{V}) and low-lying energy levels. This is limited by the use of the unique set of non-relativistic orthogonal orbitals \citep{ber95}, which is required by the ICFT $R$-matrix scattering calculation. 

%-----------------------------Table Start----------------------------
\longtab{
\begin{landscape}
\begin{longtable}{lccccccccccccc}
\caption{\label{tbl:sca_par} Thomas-Fermi-Dirac-Amaldi potential scaling parameters used in the AUTOSTRUCTURE calculations for the O-like iso-electronic sequence. $Z$ is the atomic number, e.g., 14 for silicon.}\\
\hline\hline 
\noalign{\smallskip} 
$Z$ & 1s & 2s & 2p & 3s & 3p & 3d & 4s & 4p & 4d & 4f & 5s & 5p & 5d \\ 
\noalign{\smallskip} 
\hline 
\noalign{\smallskip} 
10 & 1.45272 & 1.13455 & 1.07723 & 1.14535 & 1.11429 & 1.03711 & 1.13534 & 1.09082 & 1.10127 & 1.12000 & 1.13529 & 1.07057 & 1.07563 \\ 
\noalign{\smallskip} 
11 & 1.43928 & 1.13623 & 1.07656 & 1.15703 & 1.11601 & 1.13680 & 1.15314 & 1.09416 & 1.10714 & 1.16144 & 1.14440 & 1.08759 & 1.11304 \\ 
\noalign{\smallskip} 
12 & 1.42900 & 1.13870 & 1.07657 & 1.16882 & 1.11841 & 1.14438 & 1.15811 & 1.10176 & 1.12671 & 1.20164 & 1.13866 & 1.09188 & 1.12530 \\ 
\noalign{\smallskip} 
13 & 1.41901 & 1.14092 & 1.07641 & 1.27403 & 1.11347 & 1.25129 & 1.16709 & 1.11164 & 1.13498 & 1.25136 & 1.16597 & 1.09213 & 1.13850 \\ 
\noalign{\smallskip} 
14 & 1.41108 & 1.14281 & 1.07647 & 1.18235 & 1.12449 & 1.15380 & 1.16174 & 1.10824 & 1.14891 & 1.29649 & 1.15328 & 1.09354 & 1.11410 \\ 
\noalign{\smallskip} 
15 & 1.40435 & 1.14444 & 1.07663 & 1.17990 & 1.12631 & 1.16336 & 1.15573 & 1.11326 & 1.14965 & 1.19114 & 1.15684 & 1.10297 & 1.13895 \\ 
\noalign{\smallskip} 
16 & 1.39842 & 1.14584 & 1.07681 & 1.18274 & 1.13209 & 1.16379 & 1.15402 & 1.11215 & 1.15585 & 1.15358 & 1.16377 & 1.10945 & 1.14505 \\ 
\noalign{\smallskip} 
17 & 1.39321 & 1.14706 & 1.07701 & 1.18195 & 1.13082 & 1.20014 & 1.15741 & 1.11233 & 1.15642 & 1.08878 & 1.16759 & 1.11216 & 1.14768 \\ 
\noalign{\smallskip} 
18 & 1.38865 & 1.14813 & 1.07722 & 1.18172 & 1.13008 & 1.17355 & 1.15928 & 1.11761 & 1.15889 & 1.03165 & 1.17004 & 1.11576 & 1.14968 \\ 
\noalign{\smallskip} 
19 & 1.38468 & 1.14907 & 1.07741 & 1.18267 & 1.13006 & 1.16938 & 1.16536 & 1.11585 & 1.16002 & 1.07403 & 1.17359 & 1.11682 & 1.15303 \\ 
\noalign{\smallskip} 
20 & 1.38100 & 1.14991 & 1.07763 & 1.18304 & 1.13052 & 1.17046 & 1.17272 & 1.11924 & 1.16133 & 1.10854 & 1.18341 & 1.11859 & 1.15517 \\ 
\noalign{\smallskip} 
21 & 1.37774 & 1.15066 & 1.07785 & 1.18356 & 1.13128 & 1.17108 & 1.17177 & 1.12284 & 1.16123 & 1.12310 & 1.17128 & 1.12219 & 1.15666 \\ 
\noalign{\smallskip} 
22 & 1.37476 & 1.15134 & 1.07807 & 1.18408 & 1.13166 & 1.17223 & 1.16590 & 1.12514 & 1.16300 & 1.12925 & 1.16278 & 1.12854 & 1.15797 \\ 
\noalign{\smallskip} 
23 & 1.37211 & 1.15194 & 1.07827 & 1.18445 & 1.13203 & 1.17321 & 1.18027 & 1.12903 & 1.16483 & 1.14459 & 1.17985 & 1.12033 & 1.16132 \\ 
\noalign{\smallskip} 
24 & 1.36971 & 1.15250 & 1.07847 & 1.18493 & 1.13242 & 1.17395 & 1.17735 & 1.12907 & 1.16801 & 1.15949 & 1.16999 & 1.12260 & 1.12275 \\ 
\noalign{\smallskip} 
25 & 1.36742 & 1.15300 & 1.07866 & 1.18531 & 1.13277 & 1.17436 & 1.18096 & 1.12757 & 1.16647 & 1.15510 & 1.15632 & 1.12002 & 1.16152 \\ 
\noalign{\smallskip} 
26 & 1.36539 & 1.15346 & 1.07884 & 1.18567 & 1.13311 & 1.17486 & 1.17873 & 1.13085 & 1.16818 & 1.15826 & 1.16551 & 1.11977 & 1.16401 \\ 
\noalign{\smallskip} 
27 & 1.36354 & 1.15389 & 1.07902 & 1.18598 & 1.13343 & 1.17530 & 1.18049 & 1.12970 & 1.16746 & 1.16762 & 1.17036 & 1.12499 & 1.16463 \\ 
\noalign{\smallskip} 
28 & 1.36179 & 1.15428 & 1.07919 & 1.18628 & 1.13373 & 1.17570 & 1.18063 & 1.13108 & 1.16656 & 1.17438 & 1.16604 & 1.12448 & 1.16497 \\ 
\noalign{\smallskip} 
29 & 1.36017 & 1.15464 & 1.07935 & 1.18655 & 1.13401 & 1.17606 & 1.18062 & 1.13123 & 1.16669 & 1.17084 & 1.16777 & 1.12394 & 1.16601 \\ 
\noalign{\smallskip} 
30 & 1.35874 & 1.15498 & 1.07950 & 1.18679 & 1.13426 & 1.17638 & 1.18069 & 1.13150 & 1.16759 & 1.17330 & 1.17135 & 1.12600 & 1.16651 \\ 
\noalign{\smallskip} 
\noalign{\smallskip} 
\hline 
\end{longtable}
\end{landscape}
}
%----------------------------Table End------------------------------

\subsection{Collision}
\label{sct:col}
For the ICFT $R$-matrix collision calculation, angular momenta up to $2J=23$ and $2J=77$ are included for the exchange and non-exchange calculations, respectively. Higher angular momenta (up to infinity) are included following the top-up formula of the Burgess sum rule \citep{bur74} for dipole allowed transitions and a geometric series for the non-dipole allowed transitions \citep{bad01}. 

Three sets of outer-region $R$-matrix calculations were used for each ion. First, the resonance region is sampled with a fine energy mesh. The sampling points increase with the increasing atomic number, ranging from $\sim3600$ for \ion{Ne}{III} to $\sim30000$ for \ion{Zn}{XXIII}. Second, the energy range between the last threshold and three times the ionization potential was sampled with $\sim1000$ points for all the ions in the iso-electronic sequence. Third, a non-exchange calculation between the first threshold and three times ionization potential was performed, with $\sim1400$ sampling points for all ions in the iso-electronic sequence. Unresolved resonances in the ordinary collision strength in the resonance region are removed for the non-exchange calculation. 

By convolving the ordinary collision strength ($\Omega_{ij}$) with the Maxwellian energy distribution, we obtain the effective collision strength ($\Upsilon_{ij}$):
\begin{equation}
    \Upsilon_{ij} = \int \Omega_{ij}~\exp\left(-\frac{E}{kT}\right)~d\left(\frac{E}{kT}\right)~,
\label{eq:upsilon}
\end{equation}
where $E$ is the kinetic energy of the scattered free electron, $k$ the Boltzmann constant, and $T$ the electron temperature of the plasma. To complete the Maxwellian convolution (Equation~\ref{eq:upsilon}) at high temperatures, we calculate the infinite-energy Born and dipole line strength limits using AUTOSTRUCTURE. Between the last calculated energy point and the two limits, interpolation is used according to the type of transition in the Burgess--Tully scaled domain \citep[i.e. the quadrature of reduced collision strength over reduced energy, see][]{bur92}. 

\section{Results}
\label{sct:res}
We have obtained $R$-matrix electron-impact excitation data for the O-like iso-electronic sequence from \ion{Ne}{III} to \ion{Zn}{XXIII} (i.e., ${\rm Ne^{2+}}$ and ${\rm Zn^{22+}}$). Our effective collision strengths cover five orders of magnitude in temperature $(z+1)^2(2\times10^1,~2\times10^6)~{\rm K}$, where $z$ is the ionic charge (e.g., $z=7$ for \ion{Si}{VIII}). 

The effective collision strength data will be archived according to the Atomic Data and Analysis Structure (ADAS) data class {\it adf04} and will be available in OPEN-ADAS and our UK-APAP website\footnote{http://apap-network.org/}. These data can be used to improve the atomic database of astrophysical plasma codes, such as CHIANTI \citep{der97,dza21a} and SPEX \citep{kaa96,kaa20}, where no data or less accurate data were available. The ordinary collision strength data will also be archived in OPEN-ADAS\footnote{http://open.adas.ac.uk/}.

\section{Discussion}
\label{sct:dis}
Several ions across the iso-electronic sequence are selected to assess the quality of our structure and collision calculations. These ions were selected because detailed results from archival databases (NIST\footnote{https://www.nist.gov/pml/atomic-spectra-database} and OPEN-ADAS) and the literature are available for comparison purposes. 

\begin{figure*}
\centering
\includegraphics[width=.8\hsize, trim={1.0cm 0.5cm 1.5cm 0.5cm}, clip]{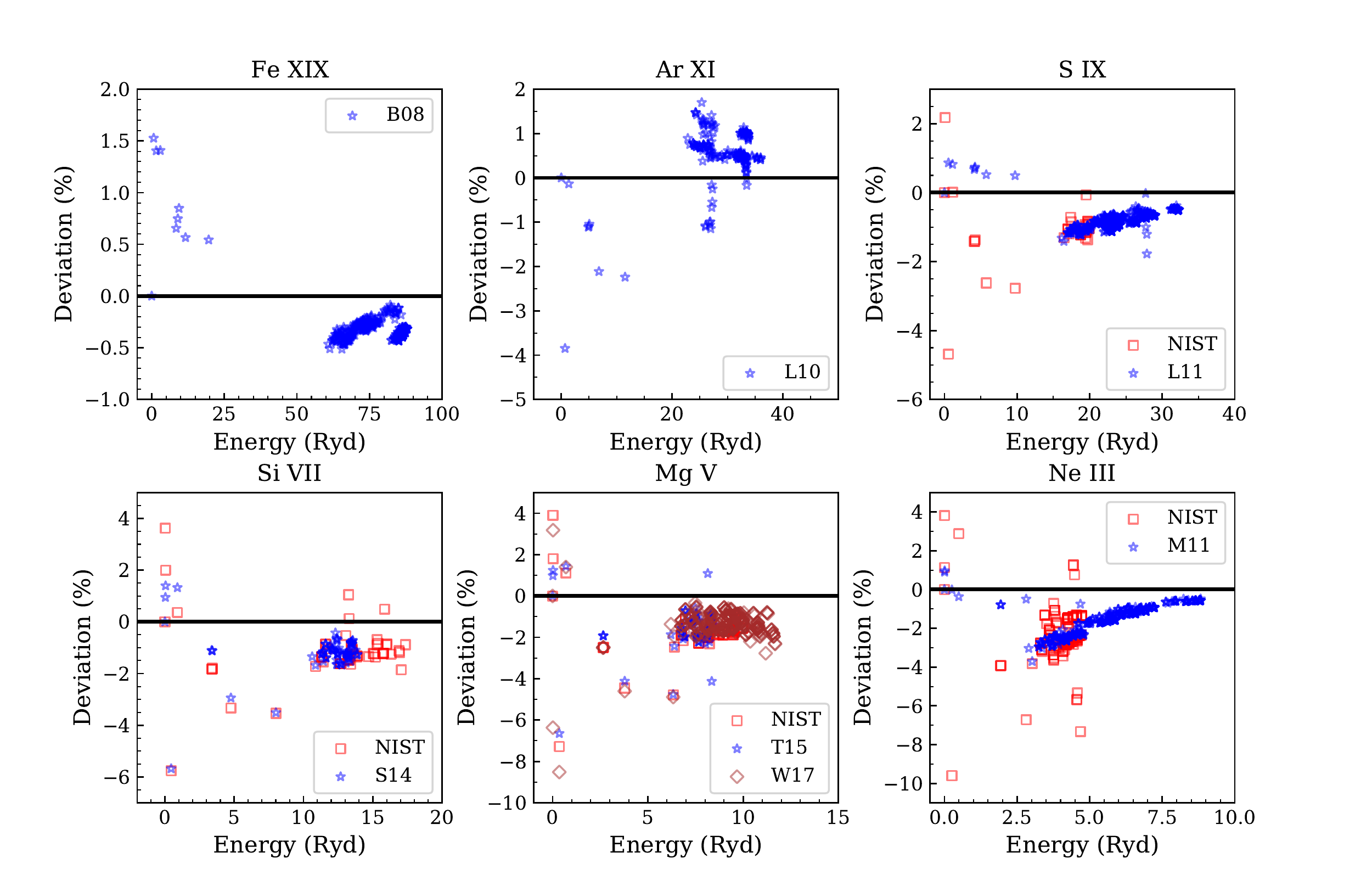}
\caption{Percentage deviations between the present energy levels (horizontal lines in black), the experimental ones (NIST) and previous works: B08 refers to \citet{but08}, L10 refers to \citet{lud10}, L11 refers to \citet{lia11}, S14 refers to \citet{sos14}, T15 refers to \citet{tay15}, W17 refers to \citet{wan17}, and M11 refers to \citet{mcl11}.}
\label{fig:plot_cflev}
\end{figure*}

First, we compare the energy levels among NIST, previous works, and the present one. As shown in Fig.~\ref{fig:plot_cflev}, generally speaking, the energy levels agree within $\sim5\%$ for the high-charge ions (e.g., \ion{Fe}{XIX} and \ion{Ar}{XI}). A larger deviation ($\lesssim10\%$) is found for low-charge ions such as \ion{Ne}{III}, in particular, for some of the low-lying energy levels.

%-----------------------------Figure Start----------------------------
%-----------------------------Figure End------------------------------

\begin{figure*}
\centering
\includegraphics[width=.8\hsize, trim={1.0cm 0.5cm 1.5cm 0.5cm}, clip]{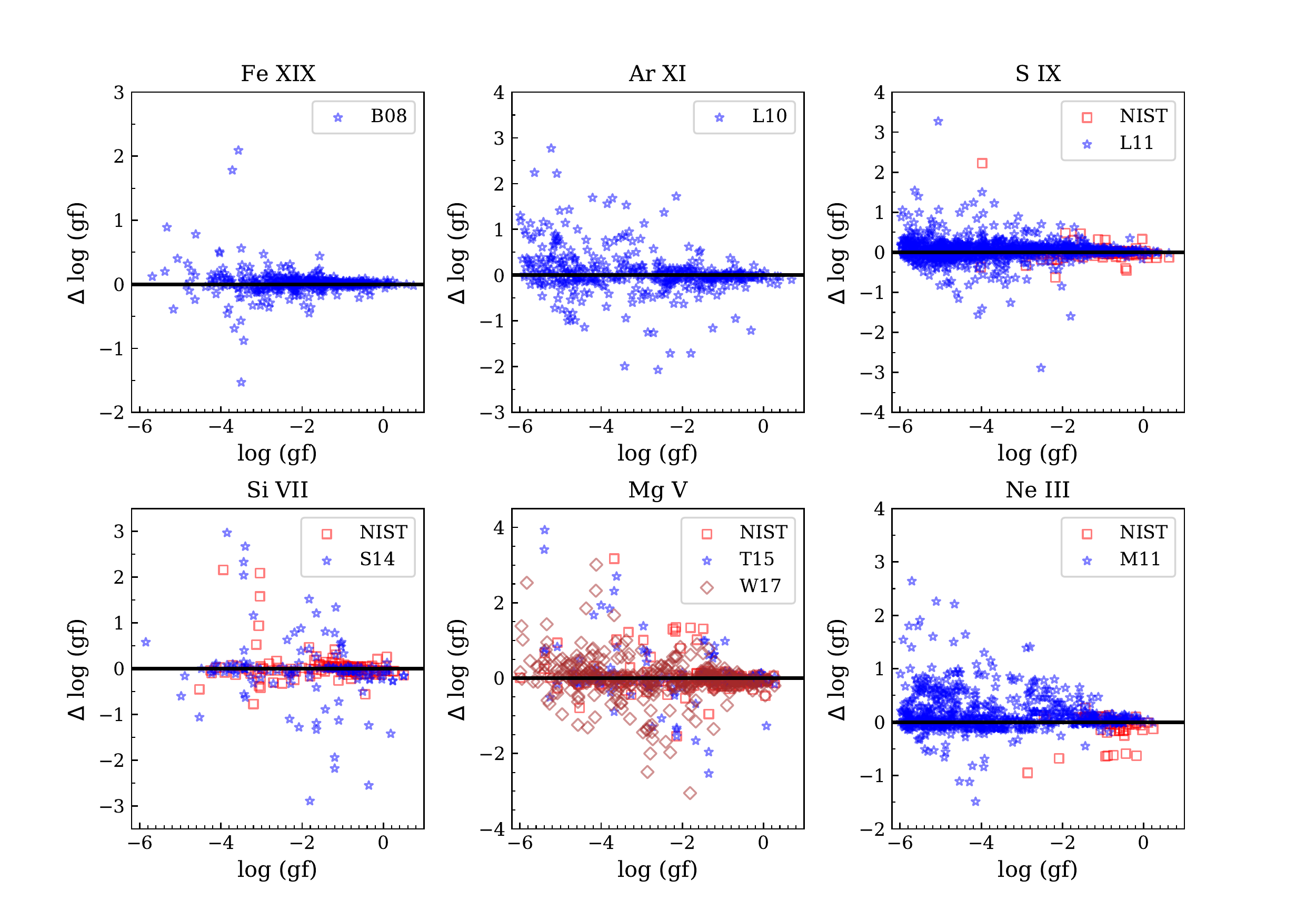}
\caption{Comparisons of $\log~(gf)$ from the present work (black horizontal line) with the experimental ones (NIST) and previous works: B08 refers to \citet{but08}, L10 refers to \citet{lud10}, L11 refers to \citet{lia11}, S14 refers to \citet{sos14}, T15 refers to \citet{tay15}, W17 refers to \citet{wan17}, and M11 refers to \citet{mcl11}. We note that this comparison is limited to relatively strong transitions with $\log~(gf) \gtrsim -6$ originating from the lowest five energy levels.}
\label{fig:plot_cftran}
\end{figure*}

Second, we compare transition strengths $\log (gf)$, where $g$ and $f$ are the statistical weight and oscillator strength of the transition, respectively. Fig.~\ref{fig:plot_cftran} shows the deviation of transition strengths $\Delta \log~(gf)$ in archival databases and previous works with respect to the present work. We limit our comparison to relatively strong transitions with $gf \gtrsim 10^{-6}$ from the lowest five energy levels of the ground configuration: $2s^22p^4 ~(^3P_{2,1,0},^1D_{2},^1S_{0})$. For low-density astrophysical plasmas, the ground and first four metastable levels \citep{mao17} dominate the level population. Weak transitions associated with higher metastable levels have little impact on astrophysical plasma diagnostics. Large difference ($\Delta \log(gf) \gtrsim 1.0$) are noticed for some transitions. Some are caused by level mixing, e.g., for Ar {\sc xi}, level \#69 $2s^2~2p^3~3d~(^1D_2)$ with $E=26.84$~Ryd and level \#87 $2s^2~2p^3~3d~(^1D_2)$ with $E=27.41$~Ryd are mixed in the present calculation. Some are associated with the last few energy levels in previous calculations, e.g., levels \#76-89 of the 92-level calculation by \citet{sos14} for Si {\sc vii} and levels \#76-86 of the 86-level calculation by \citet{tay15} for Mg {\sc v}. When comparing different size-scale structure calculations, high-lying levels can be subject to the significant effect of a different CI expansion \citep{dza15,fme16}.

%-----------------------------Figure Start----------------------------
%-----------------------------Figure End------------------------------

Hexbin plots \citep{car87} are used to compare a large number of effective collision strengths (figures in Appendix~\ref{sct:hexbin}). Statistics of the hexbin plot comparison are provided in Table~\ref{tbl:cf_ecs_stat}.
In general, at higher temperatures, different calculations agree better. At low and intermediate temperatures, the additional resonances included in the present work contribute to most of the deviations. 

\begin{table*}
\caption[]{Statistics of the hexbin plot comparison of the effective collision strength comparison for \ion{Fe}{XIX} \citep[][B08]{but08}, \ion{Ar}{XI} \citep[][L10]{lud10}, \ion{S}{IX} \citep[][L11]{lia11}, \ion{Si}{VII} \citep[][S14]{sos14}, \ion{Mg}{V} \citep{tay15,wan17}, and \ion{Ne}{III} \citep[][M11]{mcl11}. Columns \#2--\#4 give the number of transitions with $\log (\Upsilon) > -5$ in both data sets and the percentage of transitions with deviation larger than 0.2 dex at three temperatures (ion-dependent) used for the hexbin plots. Columns \#5--\#7 are the statistics when limiting the transitions from the lowest five energy levels (i.e. the ground and the first four metastable levels). For \ion{Mg}{V}, statistics of the comparisons with both \citet[][T15]{tay15} and \citet[][W17]{wan17} are shown.}
\label{tbl:cf_ecs_stat}
\centering
\begin{tabular}{c|ccc|ccc}
\hline\hline
\noalign{\smallskip} 
Ion & $T({\rm low})$ & $T({\rm middle})$ & $T({\rm high})$ & $T({\rm low})$ & $T({\rm middle})$ & $T({\rm high})$ \\
\noalign{\smallskip} 
\hline
\noalign{\smallskip} 
\ion{Fe}{XIX} (B08) & $\sim5.3\times10^4$ ($54\%$) & $\sim4.9\times10^4$ ($54\%$) & $\sim4.3\times10^4$ ($27\%$) & $1681$ ($31\%$) & $1612$ ($28\%$) & $1405$ ($6\%$)  \\
\noalign{\smallskip} 
\ion{Ar}{XI} (L10) & $\sim2.4\times10^4$ ($68\%$) & $\sim2.3\times10^4$ ($62\%$) & $\sim2.1\times10^4$ ($56\%$) & $1105$ ($44\%$) & $1091$ ($33\%$) & $1006$ ($32\%$)  \\
\noalign{\smallskip} 
\ion{S}{IX} (L11) & $4129$ ($75\%$) & $4020$ ($55\%$) & $3626$ ($34\%$) & $444$ ($29\%$) & $442$ ($31\%$) & $441$ ($23\%$)  \\
\noalign{\smallskip} 
\ion{Si}{VII} (S14) & $4092$ ($70\%$) & $4072$ ($70\%$) & $4038$ ($59\%$) & $443$ ($35\%$) & $442$ ($21\%$) & $407$ ($20\%$)  \\
\noalign{\smallskip} 
\ion{Mg}{V} (W17) & $\sim4.7\times10^4$ ($51\%$) & $\sim4.7\times10^4$ ($48\%$) & $\sim4.6\times10^4$ ($47\%$) & $1523$ ($29\%$) & $1515$ ($20\%$) & $1505$ ($22\%$)  \\
\ion{Mg}{V} (T15) & $3560$ ($62\%$) & $3547$ ($61\%$) & $3522$ ($54\%$) & $415$ ($26\%$) & $415$ ($20\%$) & $415$ ($18\%$)  \\
\noalign{\smallskip} 
\ion{Ne}{III} (M11) & $\sim1.5\times10^5$ ($83\%$) & $\sim1.5\times10^5$ ($35\%$) & $\sim1.4\times10^5$ ($25\%$) & $2435$ ($56\%$) & $2600$ ($11\%$) & $2409$ ($4\%$)  \\
\noalign{\smallskip} 
\hline
\end{tabular}
%\tablefoot{TBA}
\end{table*}

\begin{table}
\caption[]{Selected prominent transitions from the lowest four energy levels for O-like ions. The rest-frame wavelengths ($\lambda_0$ in \AA) are taken from the CHIANTI atomic database. Ground transitions are labelled with (g). Forbidden transitions are labelled with (f). }
\label{tbl:gml_tbl1}
\centering
\begin{tabular}{llll}
\hline\hline
\noalign{\smallskip} 
Ion & Lower level & Upper level & $\lambda_0$ ($\AA$) \\
\noalign{\smallskip} 
\hline
\noalign{\smallskip} 
Fe {\sc xix} & $2s^2 2p^4~(^3P_1)$ & $2s^2 2p^3 3d~(^3P_2)$ & 13.42 \\ % 3-85 in cdb
& $2s^2 2p^4~(^3P_2)$ & $2s^2 2p^3 3d~(^3D_3)$ & 13.52 (g) \\ % 1-69 in cdb
& $2s^2 2p^4~(^1D_2)$ & $2s^2 2p^3 3d~(^1F_3)$ & 13.74 \\ % 4-76 in cdb
%& $2s^2 2p^4~(^1D_2)$ & $2s 2p^5~(^1P_1)$ & 91.01 \\ % 4-9 in cdb
& $2s^2 2p^4~(^3P_2)$ & $2s 2p^5~(^3P_2)$ & 108.36 (g) \\ % 1-6 in cdb
& $2s^2 2p^4~(^3P_0)$ & $2s 2p^5~(^3P_1)$ & 109.95 \\ % 2-7 in cdb
& $2s^2 2p^4~(^3P_1)$ & $2s 2p^5~(^3P_2)$ & 119.98 \\ % 3-6 in cdb
%& $2s^2 2p^4~(^3P_1)$ & $2s^2 2p^4~(^1S_0)$ &  424.27 (f) \\ % 3-5 in cdb, note the index change
& $2s^2 2p^4~(^3P_2)$ & $2s^2 2p^4~(^1D_2)$ & 592.24 (g,~f) \\ % 1-4 in cdb, \citep{cur04}
& $2s^2 2p^4~(^3P_2)$ & $2s^2 2p^4~(^3P_1)$ & 1118.06 (g,~f) \\  % 1-3 in cdb \citep{wan06}
\noalign{\smallskip} 
\hline
\noalign{\smallskip} 
Ca {\sc xiii} & $2s^2 2p^4~(^1D_2)$ & $2s 2p^5~(^1P_1)$ & 131.22 \\ % 4-9 in cdb
& $2s^2 2p^4~(^3P_2)$ & $2s 2p^5~(^3P_2)$ & 161.74 (g) \\ % 1-6 in cdb
& $2s^2 2p^4~(^3P_1)$ & $2s^2 2p^4~(^1S_0)$ & 648.70 (f) \\ % 2-5 in cdb
& $2s^2 2p^4~(^3P_2)$ & $2s^2 2p^4~(^1D_2)$ & 1133.76 (g,~f) \\ % 1-4 in cdb
\noalign{\smallskip} 
\hline
\noalign{\smallskip} 
Ar {\sc xi} & $2s^2 2p^4~(^1D_2)$ & $2s 2p^5~(^1P_1)$ & 151.85 \\ % 4-9 in cdb
& $2s^2 2p^4~(^3P_2)$ & $2s 2p^5~(^3P_2)$ & 188.81 (g) \\ % 1-6 in cdb
%& $2s^2 2p^3 3s~(^5S_2)$ &  $2s^2 2p^3 3p~(^5P_3)$ & 717.85 \\ % 11-19
%& $2s^2 2p^3 3s~(^5S_2)$ &  $2s^2 2p^3 3p~(^5P_2)$ & 730.41 \\ % 11-18
%& $2s^2 2p^3 3s~(^5S_2)$ &  $2s^2 2p^3 3p~(^5P_1)$ & 736.42 \\ % 11-17
& $2s^2 2p^4~(^3P_1)$ & $2s^2 2p^4~(^1S_0)$ & 745.95 (f) \\ % 2-5 in cdb
& $2s^2 2p^4~(^3P_2)$ & $2s^2 2p^4~(^1D_2)$ & 1392.10 (g, f) \\ % 1-4 in cdb
\noalign{\smallskip} 
\hline
\noalign{\smallskip} 
S {\sc ix} & $2s^2 2p^4~(^3P_0)$ & $2s^2 2p^3 3d~(^3D_1)$ & 46.61 \\ % 3-83 in cdb, not observed
& $2s^2 2p^4~(^3P_2)$ & $2s^2 2p^3 3d~(^3P_2)$ & 47.25 (g) \\ % 1-69 in cdb, \citep{li13}
& $2s^2 2p^4~(^1D_2)$ & $2s^2 2p^3 3s~(^1D_2)$ & 55.54 \\ % 4-16 in cdb \citep{li13}
& $2s^2 2p^4~(^3P_1)$ & $2s^2 2p^3 3s~(^3S_1)$ & 56.33 \\ % 2-12 in cdb \citep{li13}
& $2s^2 2p^4~(^1D_2)$ & $2s 2p^5~(^1P_1)$ & 179.28 \\ % 4-9 in cdb
& $2s^2 2p^4~(^3P_2)$ & $2s 2p^5~(^3P_2)$ & 224.73 (g) \\ % 1-6 in cdb
& $2s^2 2p^4~(^3P_1)$ & $2s^2 2p^4~(^1S_0)$ & 871.73 (f) \\ % 2-5 in cdb
& $2s^2 2p^4~(^3P_2)$ & $2s^2 2p^4~(^1D_2)$ & 1715.41 (g,~f) \\ % 1-4 in cdb
\noalign{\smallskip} 
\hline
\noalign{\smallskip}
Si {\sc vii} & $2s^2 2p^4~(^1D_2)$ & $2s 2p^5~(^1P_1)$ & 217.83 \\ % 4-9 in cdb
& $2s^2 2p^4~(^3P_2)$ & $2s 2p^5~(^3P_2)$ & 275.36 (g) \\ % 1-6 in cdb
& $2s^2 2p^4~(^3P_1)$ & $2s^2 2p^4~(^1S_0)$ & 1049.15 (f) \\ % 2-5 in cdb
& $2s^2 2p^4~(^3P_2)$ & $2s^2 2p^4~(^1D_2)$ & 2147.40 (g,~f) \\ % 1-4 in cdb
\noalign{\smallskip} 
\hline
\noalign{\smallskip}
Mg {\sc v} & $2s^2 2p^4~(^1D_2)$ & $2s 2p^5~(^1P_1)$ & 276.58 \\ % 4-9 in cdb
& $2s^2 2p^4~(^3P_2)$ & $2s 2p^5~(^3P_2)$ & 353.09 (g) \\ % 1-6 in cdb
& $2s^2 2p^4~(^3P_1)$ & $2s^2 2p^4~(^1S_0)$ & 1324.43 (f) \\ % 2-5 in cdb
& $2s^2 2p^4~(^3P_2)$ & $2s^2 2p^4~(^1D_2)$ & 2783.58 (g,~f) \\ % 1-4 in cdb
\noalign{\smallskip} 
\hline
\noalign{\smallskip}
Ne {\sc iii} & $2s^2 2p^4~(^1D_2)$ & $2s 2p^5~(^1P_1)$ & 379.31 \\ % 4-11 in cdb, note the index change
& $2s^2 2p^4~(^3P_2)$ & $2s 2p^5~(^3P_2)$ & 489.50 (g) \\ % 1-6 in cdb
& $2s^2 2p^4~(^3P_1)$ & $2s^2 2p^4~(^1S_0)$ & 1814.63 (f) \\ % 2-5 in cdb
& $2s^2 2p^4~(^3P_2)$ & $2s^2 2p^4~(^1D_2)$ & 3869.85 (g,~f) \\ % 1-4 in cdb
\noalign{\smallskip}
\hline
\end{tabular}
%\tablefoot{TBA}
\end{table}

In the following, we compare selected prominent allowed and forbidden transitions (Table~\ref{tbl:gml_tbl1}) from the ground and metastable levels for \ion{Fe}{XIX} (Section~\ref{sct:082619}), \ion{Ca}{XIII} (Section~\ref{sct:082013}), \ion{Ar}{XI} (Section~\ref{sct:081811}), \ion{S}{IX} (Section~\ref{sct:081609}), \ion{Si}{VII} (Section~\ref{sct:081407}), \ion{Mg}{V} (Section~\ref{sct:081205}), and \ion{Ne}{III} (Section~\ref{sct:081003}). 

\subsection{\ion{Fe}{XIX}}
\label{sct:082619}
The most recent calculation of $R$-matrix electron-impact excitation data for \ion{Fe}{XIX} (or ${\rm Fe^{18+}}$) is presented by \citet[][B08 hereafter]{but08}. Both B08 and the present work use AUTOSTRUCTURE for the structure calculation. As shown in the top-left panels of Fig.~\ref{fig:plot_cflev} and Fig.~\ref{fig:plot_cftran}, the energy levels and transition strengths agree well between the present work and B08. 

\begin{figure}
\centering
\includegraphics[width=\hsize, trim={0.cm 0.5cm 1.0cm 0.5cm}, clip]{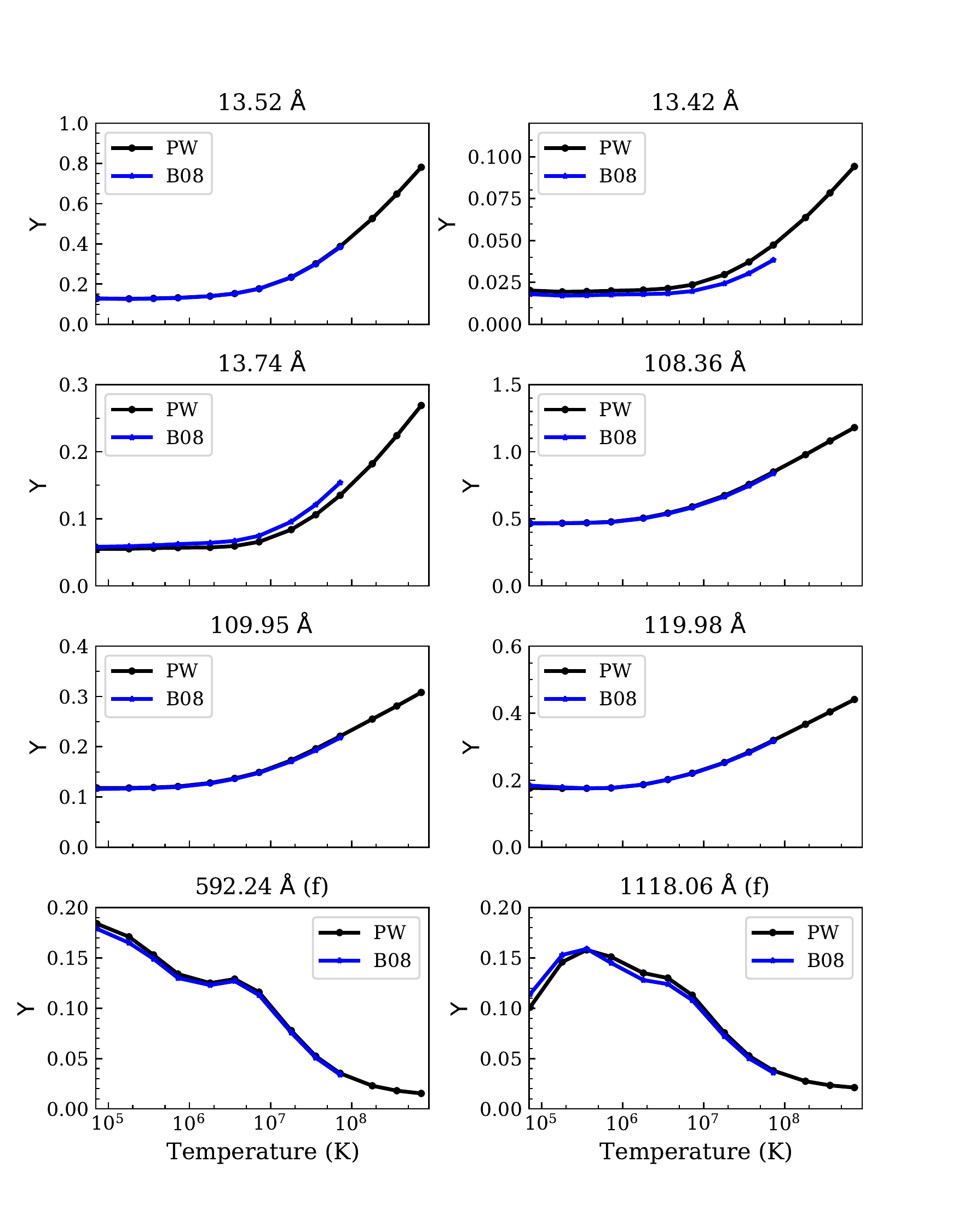}
\caption{Comparison of \ion{Fe}{XIX} (or ${\rm Fe^{18+}}$) effective collision strengths between the present work (PW) and \citet[][B08]{but08} for selected dipole transitions from the ground (upper) and metastable (middle and bottom) levels listed in Table~\ref{tbl:gml_tbl1}. }
\label{fig:plot_ecs_082619}
\end{figure}

Both B08 and the present work use the $R$-matrix ICFT method for the scattering calculation. B08 included 342 fine-structure levels in the close-coupling expansions, which is smaller than the present work (630 levels). Fig.~\ref{fig:hexbin_cfecs_082619} shows the hexbin plot comparison of the effective collision strengths at $T=1.80\times10^5~{\rm K}$, $3.61\times10^6~{\rm K}$, and $7.22\times10^7~{\rm K}$. The effective collision strengths for the six selected dipole transitions from the ground (13.52~\AA\ and 108.36~\AA) and metastable (13.42~\AA, 13.74~\AA, 109.95~\AA, and 119.98~\AA) levels, as well as two forbidden transitions (592.24~\AA\ and 1118.06~\AA) agree well between the present work and B08 (Fig.~\ref{fig:plot_ecs_082619}).

%-----------------------------Figure Start----------------------------
%-----------------------------Figure End------------------------------

\subsection{\ion{Ca}{XIII}}
\label{sct:082013}
\begin{figure}
\centering
\includegraphics[width=.87\hsize, trim={0.cm 0.5cm 0.5cm 0.5cm}, clip]{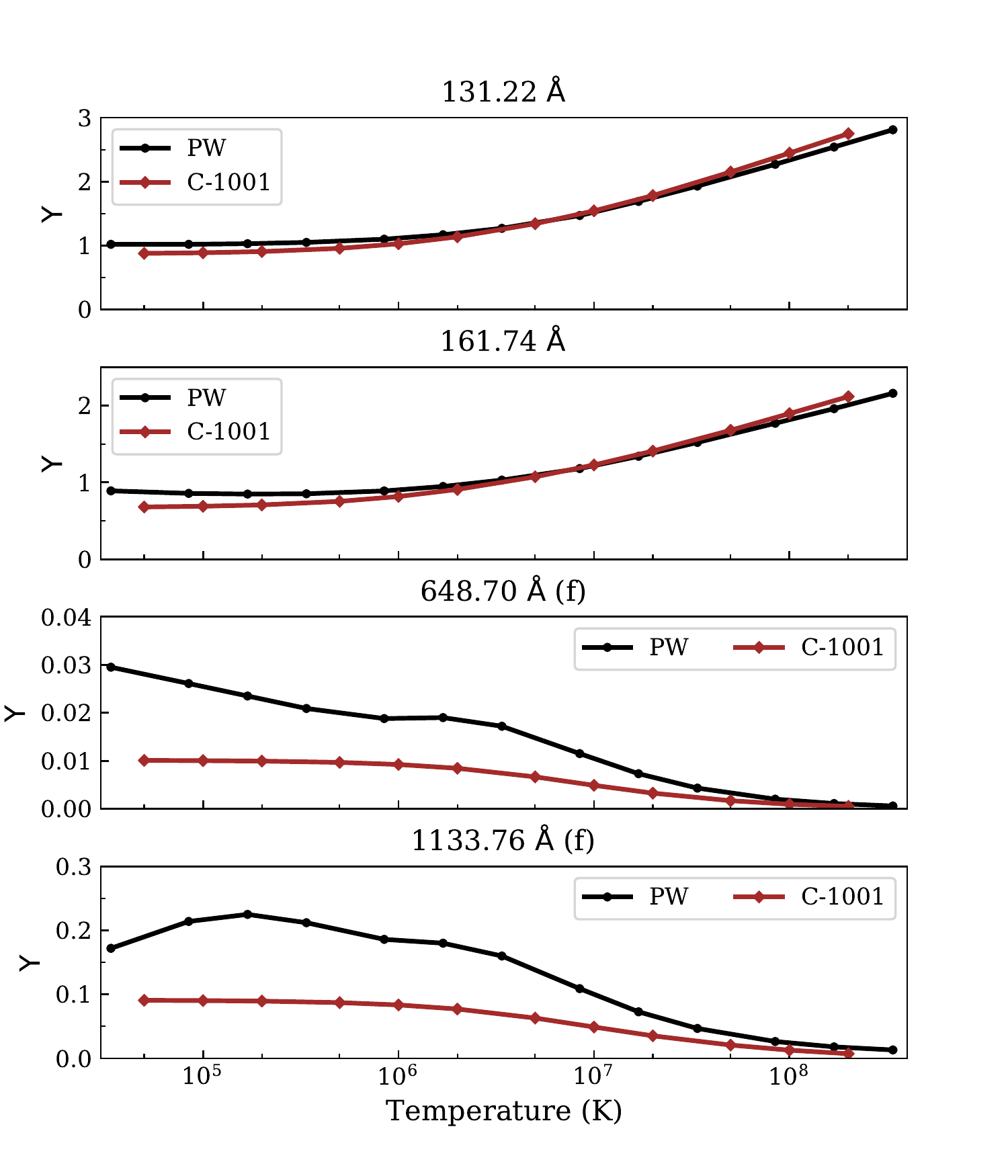}
\caption{Comparison of \ion{Ca}{XIII} (or ${\rm Ca^{12+}}$) effective collision strengths between the present work (PW), and \citet[][L05]{lan05ca} as included in the CHIANTI atomic database v10.0.1 (C-1001). The top two panels are allowed transitions from the ground (131.22~\AA) and metastable (161.74~\AA) levels. The bottom two panels are forbidden transitions (648.70~\AA\ and 1133.76~\AA). }
\label{fig:plot_ecs_082013}
\end{figure}

The collision data of \ion{Ca}{XIII} in the CHIANTI database \citep[v10][]{dza21a} originate from \citet{lan05ca}, which presents a distorted wave calculation of 86 levels. Figure~\ref{fig:plot_ecs_082013} compares the effective collision strength of selected allowed and forbidden transitions listed in Table~\ref{tbl:gml_tbl1}. The two calculations agree well for the allowed transitions but differ for the forbidden transitions. This is similar to the case of C-like \ion{Ar}{XIII} discussed in \citet{mao20c}. The present rates result in a significant increase in the predicted line intensities of the UV forbidden lines, which have been observed by e.g., SOHO/SUMER.

%-----------------------------Figure Start----------------------------
%-----------------------------Figure End------------------------------

\subsection{\ion{Ar}{XI}}
\label{sct:081811}
\citet[][L10 hereafter]{lud10} presented the most recent $R$-matrix calculations of electron-impact excitation data for \ion{Ar}{XI} (or ${\rm Ar^{10+}}$). Both L10 and the present work use AUTOSTRUCTURE for the structure calculation. The energy levels of the L10 calculations are within $\sim2-4~\%$ of the present work (Fig.~\ref{fig:plot_cflev}). The transition strengths agree well between L10 and the present work.

\begin{figure}
\centering
\includegraphics[width=.87\hsize, trim={0.cm 0.5cm 0.5cm 1.2cm}, clip]{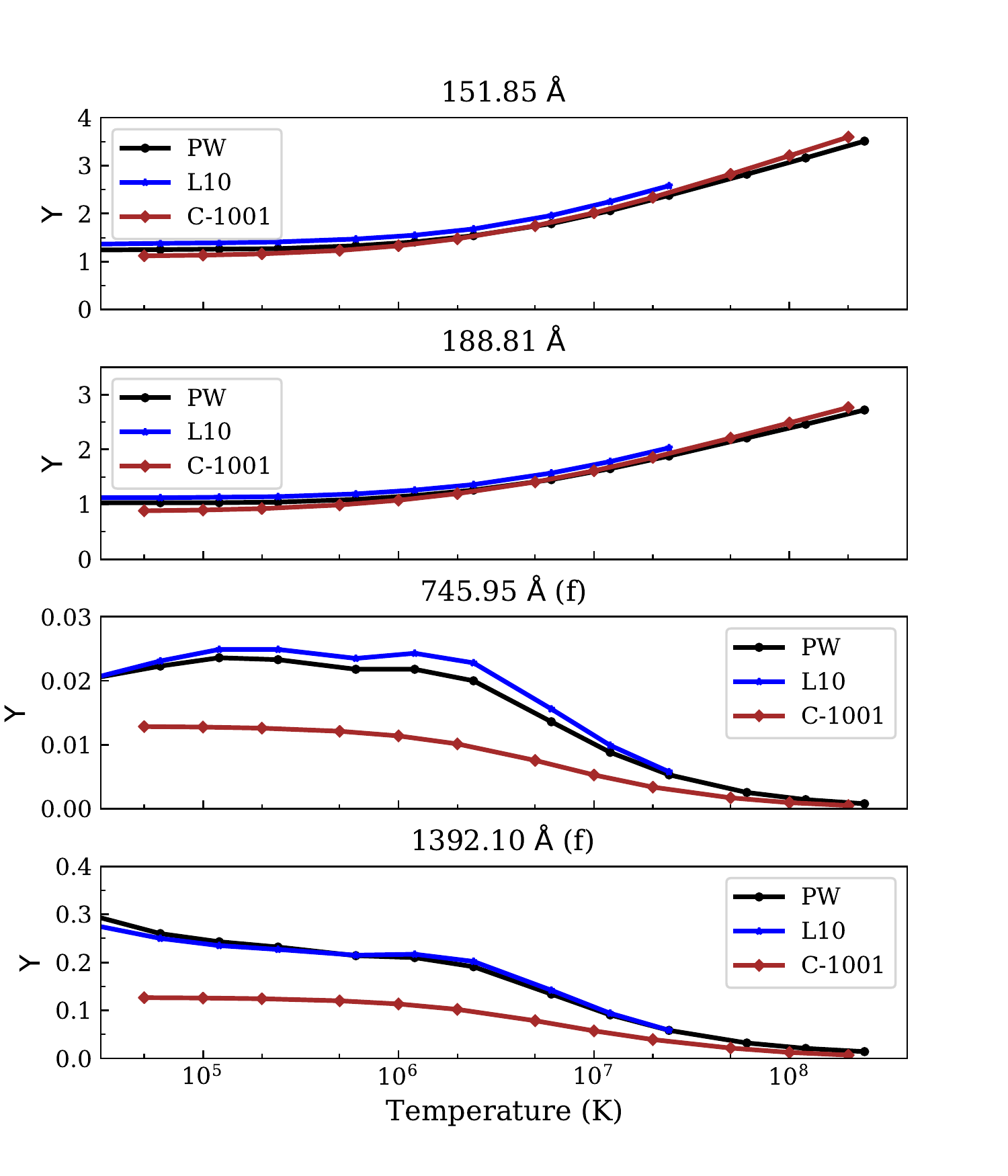}
\caption{Comparison of \ion{Ar}{XI} (or ${\rm Ar^{10+}}$) effective collision strengths between the present work (M20), \citet[][L10]{lud10}, and \citet[][distorted wave]{lan06} as incorporated in the CHIANTI atomic database v10.0.1 (C-1001). The top two panels are allowed transitions from the ground (188.81~\AA) and metastable (151.85~\AA) levels. The bottom two panels are forbidden transitions (745.95~\AA\ and 1392.10~\AA).  }
\label{fig:plot_ecs_081811}
\end{figure}

The $R$-matrix ICFT method is used for the scattering calculation of L10 (228 levels) and the present work (630 levels). Fig.~\ref{fig:hexbin_cfecs_081811} shows the hexbin plot comparison of the effective collision strengths at three temperatures in the range of $10^{5-7}$~K. Fig.~\ref{fig:plot_ecs_081811} shows the comparison of the effective collision strengths of selected allowed and forbidden transitions from the ground and metastable levels listed in Table~\ref{tbl:gml_tbl1}. The $R$-matrix calculations (L10 and the present work) agree well for all the highlighted transitions. For the forbidden transitions, the distorted wave (DW) collision strengths \citep{lan06} as available in the CHIANTI atomic database v10.0.1 \citep{dza21a} are systematically lower. 

%-----------------------------Figure Start----------------------------
%-----------------------------Figure End------------------------------

The resonant dipole-allowed transition, at 188.81~\AA, is observed by the Hinode/EIS spectrometer, while the forbidden transitions listed in Table~\ref{tbl:gml_tbl1}, as well as several allowed transitions from the 2s$^2$ 2p$^3$ 3p to 2s$^2$ 2p$^3$ 3s states have been observed by the SOHO/SUMER instrument in the 715--740~\AA\ range. These latter transitions, together with the forbidden transition at 745.80~\AA, are observable by the latest solar UV spectrometer, SPICE \citep{spc20}, on board the ESA mission Solar Orbiter, launched in 2020. These transitions are particularly useful as are very close in wavelength and could be used to measure the electron temperature, as we show here for the first time.

\begin{figure}
\centering
\includegraphics[width=.98\hsize, trim={1.5cm 1.5cm 0.5cm 1.5cm}, clip]{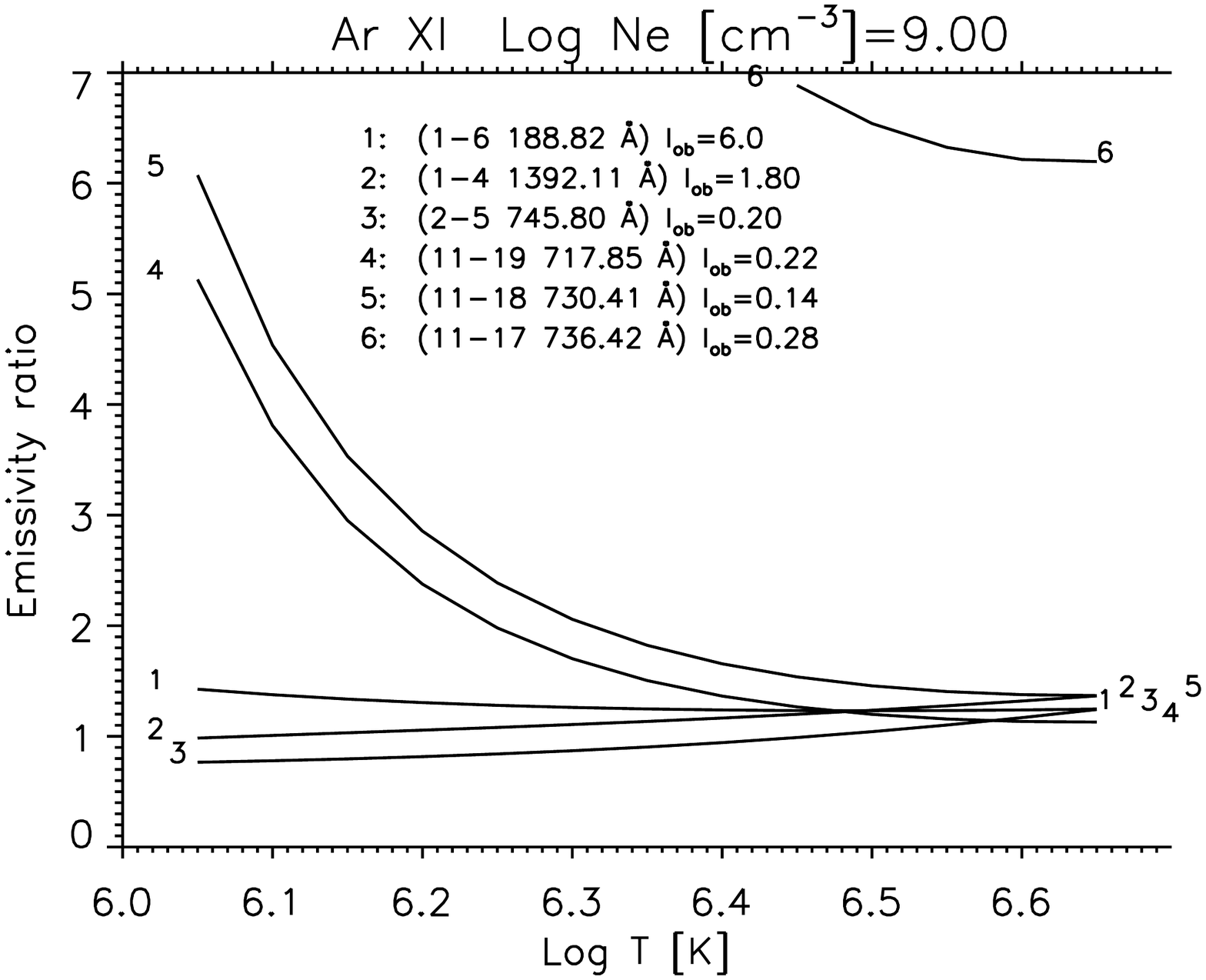}
\includegraphics[width=.98\hsize, trim={1.5cm 1.5cm 0.5cm 1.5cm}, clip]{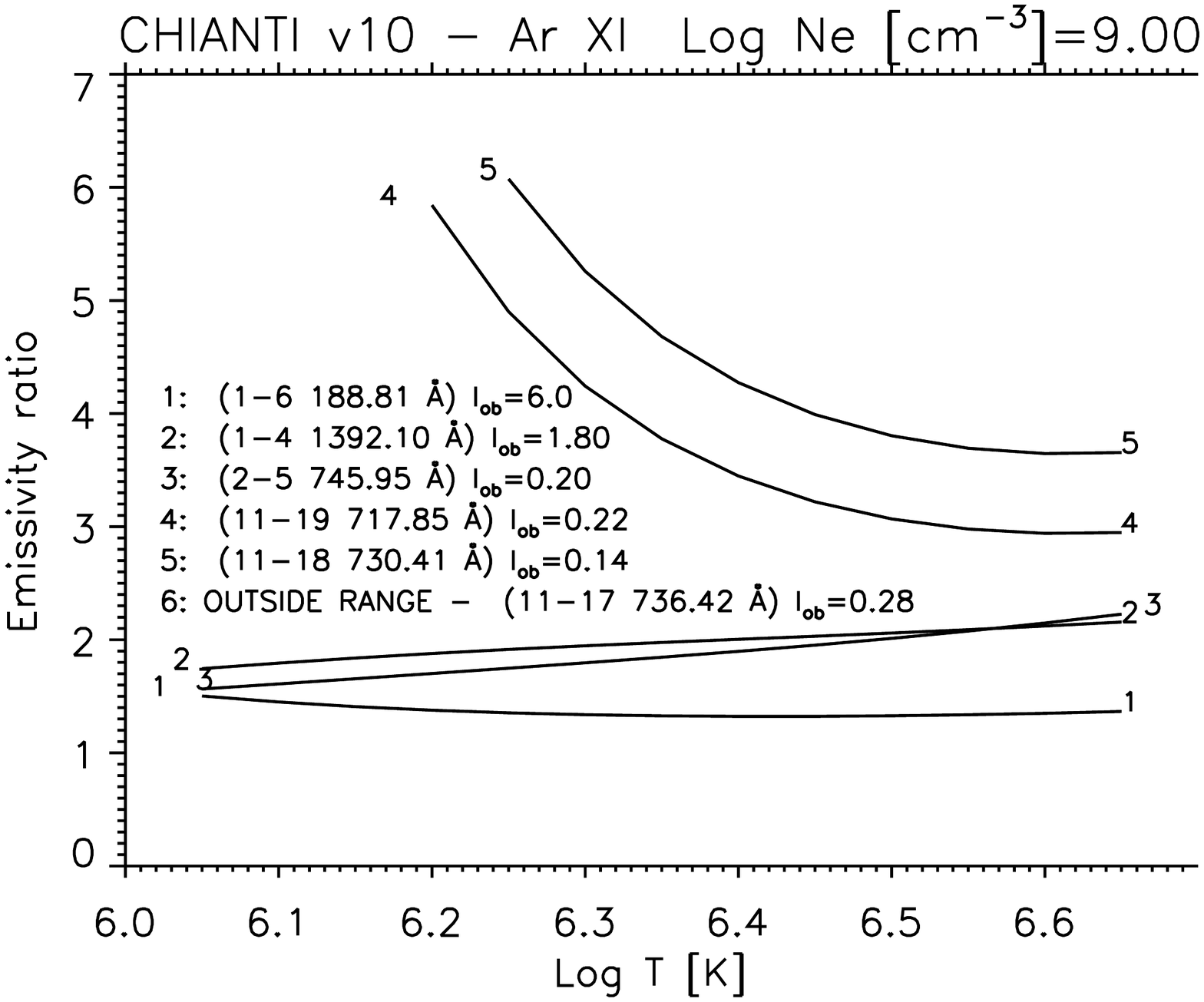}
\caption{Emissivity ratios of \ion{Ar}{XI} transitions, calculated with the present atomic rates (top) and CHIANTI atomic data (bottom). The values in brackets indicate the lower and upper level number and the wavelength (\AA). I$_{\rm ob}$ indicates the observed radiances (photon units) in active regions.}
\label{fig:ar_11_emratios}
\end{figure}

We have considered a SUMER off-limb coronal observation above an active region reported by \cite{cur04}, and an Hinode/EIS measurement of the resonant transition (noting that the SUMER and EIS observations were not simultaneous). Fig.~\ref{fig:ar_11_emratios} (top) shows the `emissivity ratios' of these lines, using the present atomic data. The emissivity ratios are essentially the ratios of the observed radiances in the lines with their emissivities, as a function of the electron temperature for a fixed density, scaled by an arbitrary constant to make the ratios close to unity \citep[see ][for details and examples]{dza04,dza18a}. If a temperature exists for which the line radiances agree with theory, the emissivity ratios would show a crossing. 

The emissivities of the chosen lines have a minor dependence on the electron density. We have taken a typical value for an active region, $10^{9}~{\rm cm^{-3}}$. To within 20\%, there is excellent agreement between predicted and observed emissivities for a typical temperature of an active region $\sim3\times10^6$~K. The only exception is the 736.42~\AA\ line, which is about 6 times stronger than predicted. This line was listed as the strongest Ar {\sc xi} line in \citet{cur04}, but the present atomic data clearly indicate that the 736.42~\AA\ line must be mostly due to another transition. the line is not due to \ion{Ar}{XI}. The 20\% scatter could be due to uncertainties in the measurements or calibration, and could be further improved with more accurate radiative data. 

Fig.~\ref{fig:ar_11_emratios} (bottom) shows instead the emissivity ratios of the same lines, using the CHIANTI atomic data. The emissivity ratio of the resonance line at 188.81~\AA\ is nearly the same, while those of the forbidden lines are significantly higher, reflecting the fact that the emissivities of the lines are much lower, mainly due to the lower DW excitation rates as shown in Fig.~\ref{fig:plot_ecs_081811}. The emissivities of the 3s--3p transitions (11--17, 11--18, 11--19 are underestimated by large factors, over three.

%-----------------------------Figure Start----------------------------
%-----------------------------Figure End------------------------------

\subsection{\ion{S}{IX}}
\label{sct:081609}
The most recent $R$-matrix calculations of electron-impact excitation data for \ion{S}{IX} (or ${\rm S^{8+}}$) are presented in \citet[][L11 hereafter]{lia11}. Both L11 and the present work all used AUTOSTRUCTURE for the atomic structure calculation. As shown in the top-right panels of Fig.~\ref{fig:plot_cflev} and Fig.~\ref{fig:plot_cftran}, the energy levels and transition strengths agree well between the present work and L11. 

\begin{figure}
\centering
\includegraphics[width=\hsize, trim={0.cm 0.5cm 0.5cm 0.5cm}, clip]{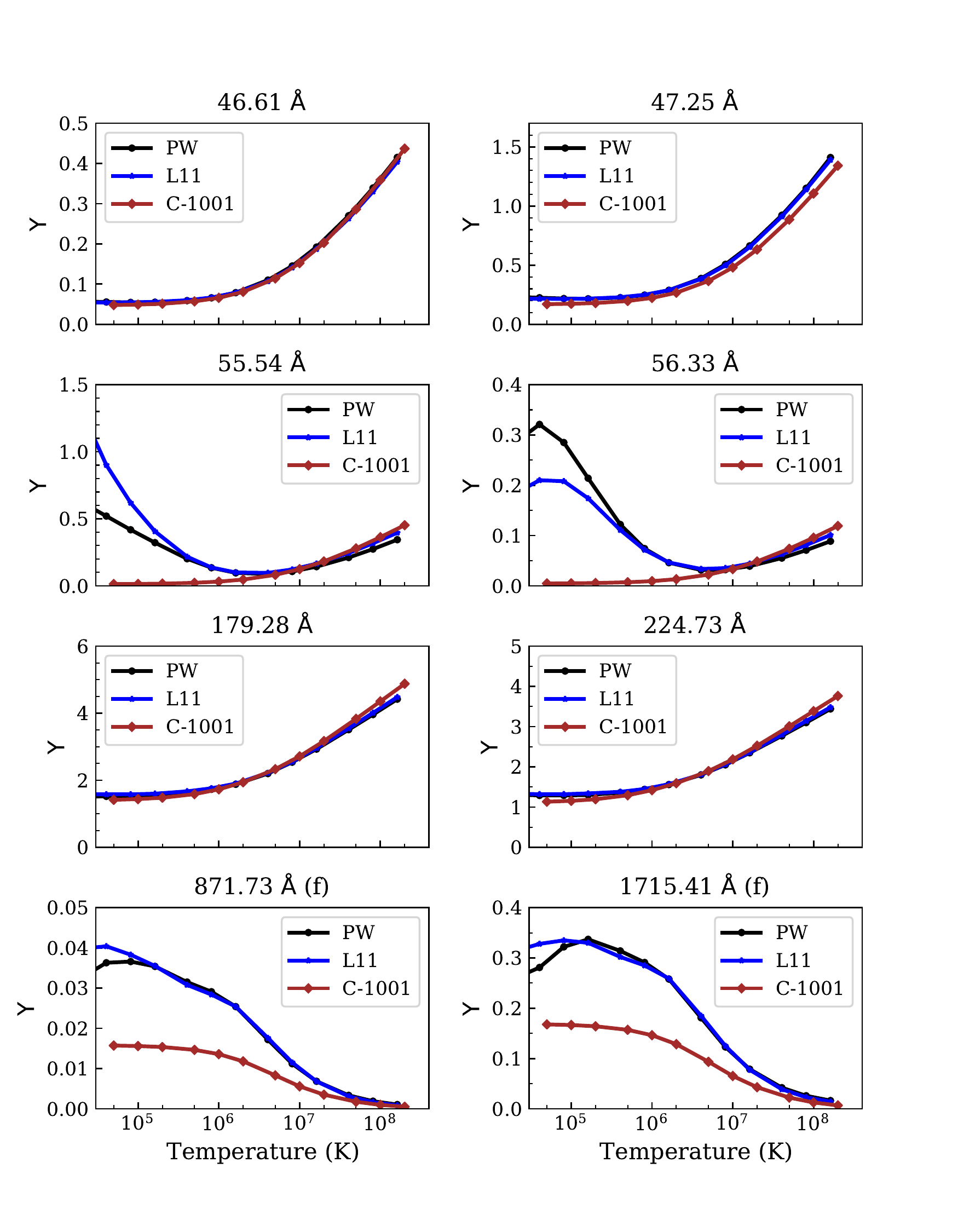}
\caption{Comparison of \ion{S}{IX} (or ${\rm S^{8+}}$) effective collision strengths between the present work (PW), \citet[][L11]{lia11}, and \citet[][distorted wave]{bha03s} as incorporated in the CHIANTI atomic database v10.0.1 (C-1001) for selected transitions listed in Table~\ref{tbl:gml_tbl1}. }
\label{fig:plot_ecs_081609}
\end{figure}

The $R$-matrix ICFT method is used for the scattering calculation of L11 (92 levels) and the present work (630 levels). Fig.~\ref{fig:hexbin_cfecs_081609} shows the hexbin plot comparison of the effective collision strengths at three temperatures in the range of $10^{4-8}$~K. As shown in Fig.~\ref{fig:plot_ecs_081609}, the effective collision strengths of three selected dipole transitions from the ground and metastable levels agree well between the four data sets: present work (PW), \citet[][L11]{lia11}, and \citet[][distorted wave]{bha03s} as incorporated in the CHIANTI atomic database v10.0.1 (C-1001). All three calculations agree well for the transitions of $46.61$~\AA\ (metastable), $47.25$~\AA\ (ground),
$179.28$~\AA\ (metastable), and $224.73$~\AA\ (ground). For the two  transitions $55.54$~\AA\ and $56.33$~\AA, the three calculations differ at $T\lesssim10^{6}$~K. For the two forbidden transitions $871.73$~\AA\ and $1715.41$~\AA, the $R$-matrix data only differ at $T\lesssim10^5$~K, while the distorted wave data is systematically smaller by a factor of $\lesssim2$. 

%-----------------------------Figure Start----------------------------
%-----------------------------Figure End------------------------------

\begin{figure}
\centering
\includegraphics[width=.98\hsize, trim={1.5cm 1.5cm 0.5cm 1.5cm}, clip]{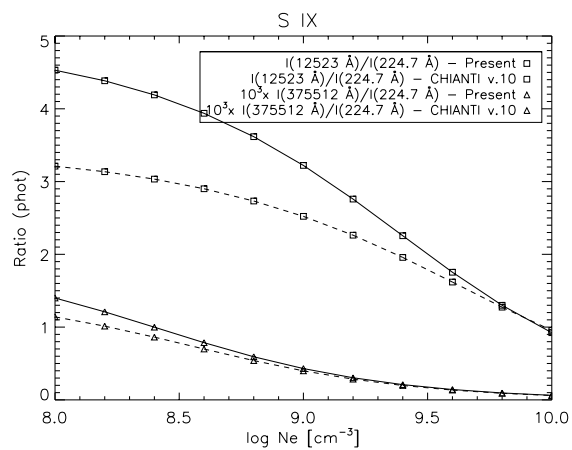}
\caption{Ratios of the two main \ion{S}{IX} near-infrared transitions, calculated with the present atomic rates (full lines) and CHIANTI atomic data (dashed lines), relative to the resonance line at $224.7$~\AA.}
\label{fig:s_9_ratios}
\end{figure}

\begin{figure}
\centering
\includegraphics[width=.98\hsize, trim={1.5cm 1.5cm 0.5cm 1.5cm}, clip]{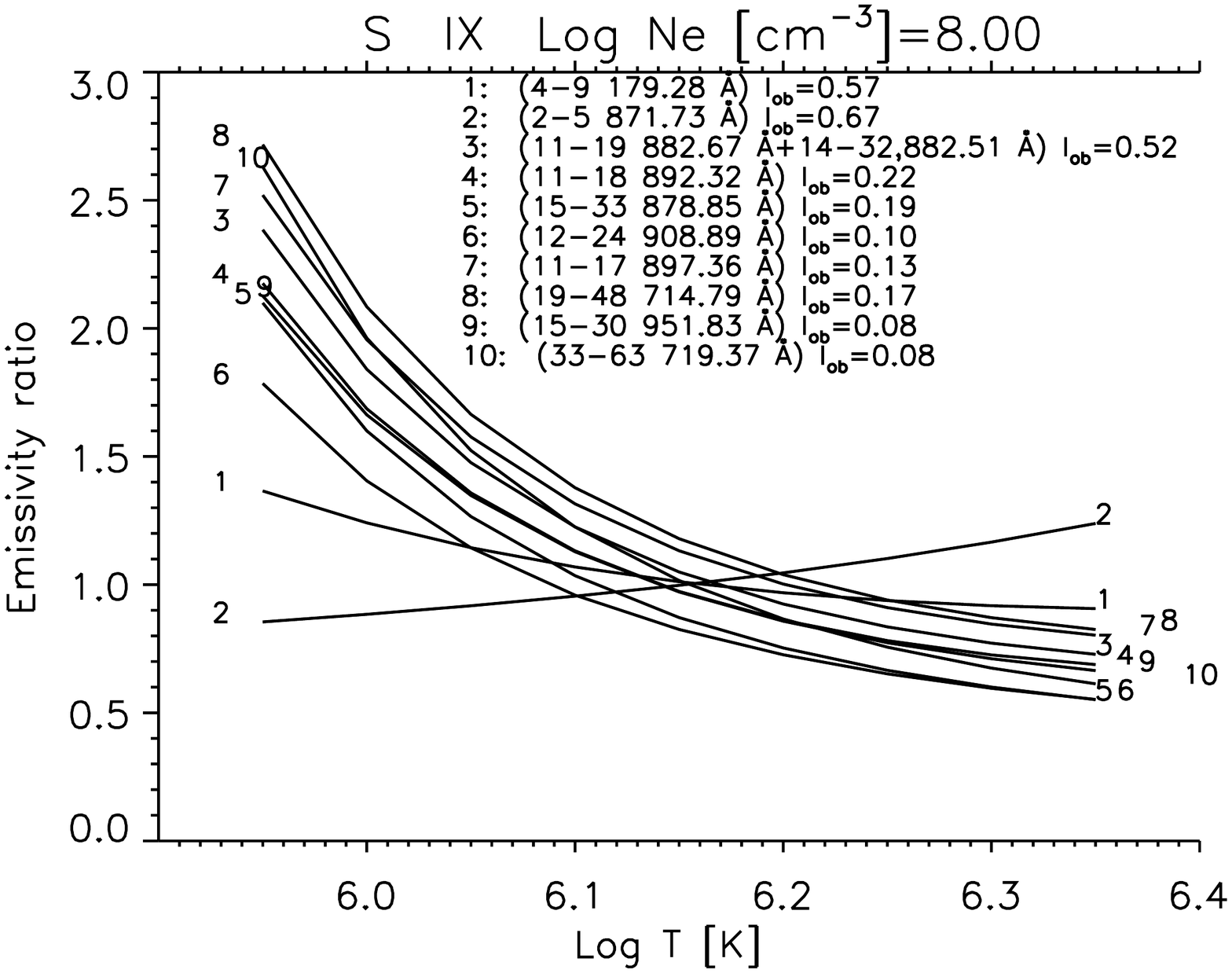}
\includegraphics[width=.98\hsize, trim={1.5cm 1.5cm 0.5cm 1.5cm}, clip]{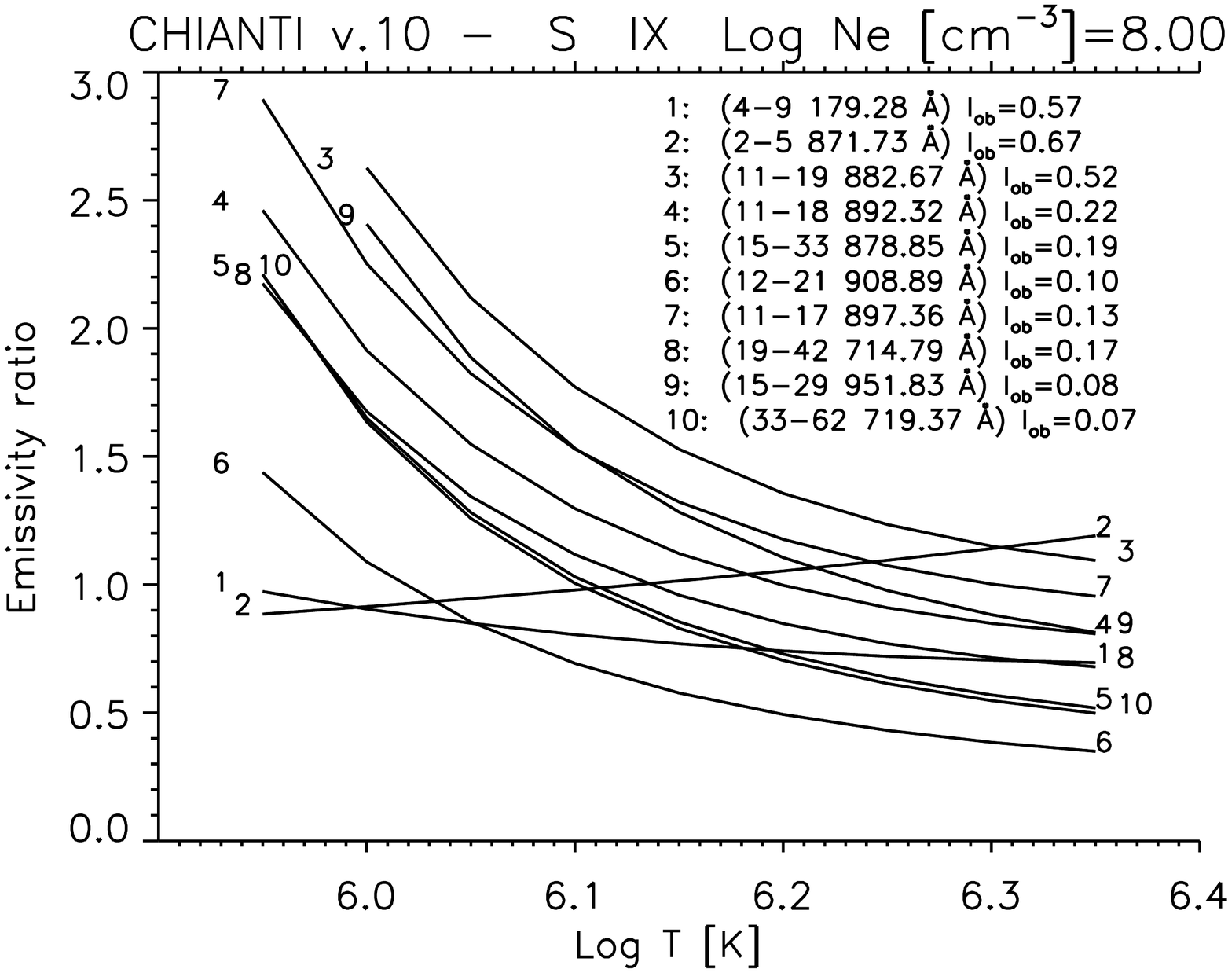}
\caption{Emissivity ratios of \ion{S}{IX} transitions, calculated with the present atomic rates (top) and CHIANTI atomic data (bottom). The values in brackets indicate the lower and upper level number and the wavelength (\AA). I$_{\rm ob}$ indicates the observed radiances (photon units) in the quiet Sun.}
\label{fig:s_9_emratios}
\end{figure}

Within the ground configuration ($2s^2 2p^4$), the main transition is the $^3P_2$--$^3P_1$ in the near-infrared (NIR) at 1.25 $\mu$\,m. It is one of the primary lines for the first large-scale (4 meter) solar ground-based telescope,  the Daniel K. Inouye Solar Telescope (DKIST, see \citealt{rim15}) and its main NIR instrument, the CryoNIRSP spectropolarimeter \citep{feh16} to measure the sulphur abundance, as this is one of the very few NIR lines to have been already observed (for a discussion on the diagnostics lines in the NIR see \citealt{dza18b}). The weaker $^3P_1$--$^3P_0$ transition at 3.75 $\mu$\,m is also observable by DKIST. Fig.~\ref{fig:s_9_ratios} shows that with the present atomic data the intensity of the 1.25 $\mu$\,m transition is up to 40\% higher than what is calculated by CHIANTI with the DW collisional rates. Note that the forbidden lines in the off-limb observations can be strongly influenced by photoexcitation (PE) from the disk radiation, depending on the distance from the photosphere and the local electron density. PE is not included in Fig.~\ref{fig:s_9_ratios}, but is included in Fig.~\ref{fig:s_9_emratios}, where we plot the emissivity ratios of several UV transitions, relative to the SUMER off-limb quiet Sun coronal observation reported by \cite{cur04}. We have added a measurement from Hinode/EIS of the EUV line at 179.28~\AA. 
 
As in the \ion{Ar}{XI} case, the transitions shown in Fig.~\ref{fig:s_9_emratios} provide a very good temperature diagnostic, as they are close in wavelength. With the present atomic data (top plot), despite the relatively large scatter (about 20\%, mostly in the weaker lines), the curves indicate an electron temperature of $1.4\times10^6$~K, in excellent agreement with a few other quiet Sun measurements as discussed in \cite{dza18a}. On the other hand, using the CHIANTI atomic data, a large scatter (factor of two) in the curves is present, as shown in Fig.~\ref{fig:s_9_emratios} (bottom), and an incorrect temperature would have been estimated. 

Finally, we note that a few of the weaker \ion{S}{IX} transitions around 715~\AA\ are also observable by the Solar Orbiter SPICE spectrometer, although we point out that several identifications reported by \cite{cur04} need to be revised in light of the present atomic data.

%-----------------------------Figure Start----------------------------
%-----------------------------Figure End------------------------------

\subsection{\ion{Si}{VII}}
\label{sct:081407}
The most recent $R$-matrix calculations of electron-impact excitation data for \ion{Si}{VII} (or ${\rm Si^{6+}}$) are presented in \citet[][S14]{sos14}. S14 used the multi-configuration Hartree-Fock method, which is different from the present work. As shown in the bottom-left panel of Fig.~\ref{fig:plot_cflev}, the level energies of NIST and S14 agree well, while the energy levels of the present work differ up to $\sim6~\%$. The transition strengths of NIST, S14, and the present work agree well with each other (the bottom-left panel of Fig.~\ref{fig:plot_cftran}). 

\begin{figure}
\centering
\includegraphics[width=.87\hsize, trim={0.cm 0.5cm 0.5cm 0.5cm}, clip]{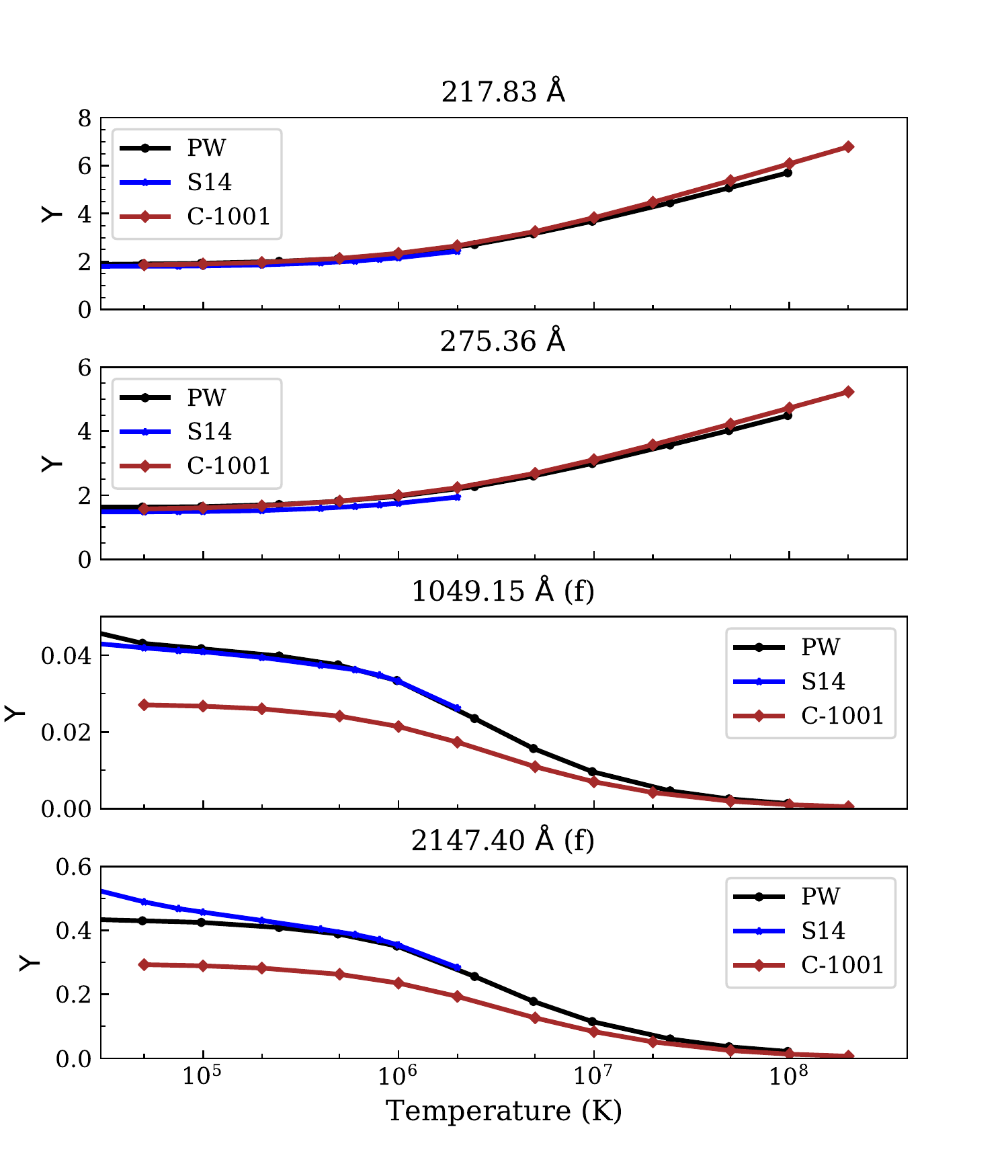}
\caption{Comparison of \ion{Si}{VII} (or ${\rm Si^{6+}}$) effective collision strengths between the present work (PW), \citet[][S14]{sos14}, and \citet[][distorted wave]{bha03si} as incorporated in the CHIANTI atomic database v10.0.1 (C-1001) for selected transitions listed in Table~\ref{tbl:gml_tbl1}. }
\label{fig:plot_ecs_081407}
\end{figure}

S14 used the B-spline $R$-matrix method for the scattering calculation, including 92 fine-structure target levels. Fig.~\ref{fig:hexbin_cfecs_081407} shows the hexbin plot comparison of the effective collision strengths at three temperatures in the range of $10^{4-6}$~K. In Fig.~\ref{fig:plot_ecs_081407}, we compare the effective collision strengths of selected transitions listed in Table~\ref{tbl:gml_tbl1}.
We also consider the \citet[][distorted wave]{bha03si} data as available in CHIANTI. Similar to \ion{Ar}{XI}, the three calculations agree well for the allowed transitions, but for the forbidden transitions, the distorted wave rates are systematically lower than those of the $R$-matrix calculations.

The present rates result in a significant increase in the predicted intensity of the 1049.15~\AA\ transition,
compared to what is calculated within CHIANTI. We note that the 1049.15~\AA\ transition is one of the primary lines for the Solar Orbiter SPICE instrument, to measure chemical abundance variations. In fact, most lines observable by SPICE are due to elements with a high first ionization potential (FIP) such as Ne, Ar, O, and very few transitions from low-FIP elements such as Si are available. Such measurements are important as the chemical abundances of low-FIP vs. high-FIP elements in the solar corona and solar wind vary \citep[see the review in ][]{dza18a}.

%-----------------------------Figure Start----------------------------
%-----------------------------Figure End------------------------------

\subsection{\ion{Mg}{V}}
\label{sct:081205}
The most recent $R$-matrix calculations of  electron-impact excitation data for \ion{Mg}{V} (or ${\rm Mg^{4+}}$) are presented in \citet[][T15]{tay15} and \citet[][W17]{wan17}. Both T15 and W17 used the multi-configuration Hartree-Fock method. As shown in the bottom-middle panel of Fig.~\ref{fig:plot_cflev}, the level energies of NIST, T15, and W17 agree well among each other, while the energy levels of the present work differ by up to $\sim9~\%$. The transition strengths of NIST, T15, W17, and the present work agree well with each other (see the bottom-left panel of Fig.~\ref{fig:plot_cftran}). 

\begin{figure}
\centering
\includegraphics[width=.87\hsize, trim={0.cm 0.5cm 0.5cm 0.5cm}, clip]{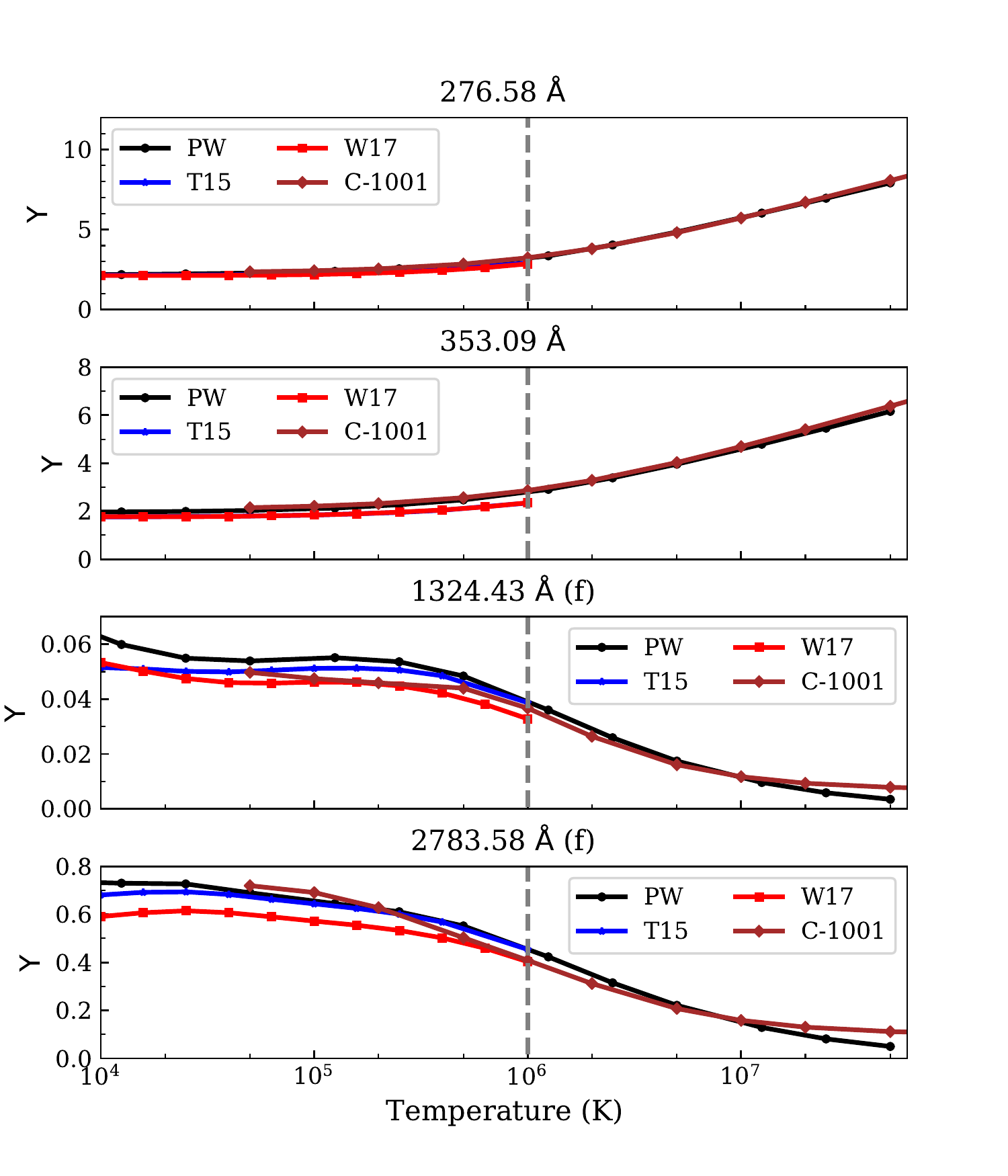}
\caption{Comparison of \ion{Mg}{V} (or ${\rm Mg^{4+}}$) effective collision strengths between the present work (PW), \citet[][T15]{tay15}, \citet[][W17]{wan17}, and composite data as incorporated in the CHIANTI atomic database v10.0.1 (C-1001) for selected transitions listed in Table~\ref{tbl:gml_tbl1}. The vertical dashed lines indicate the temperature threshold ($T\sim10^5$~K), below which $R$-matrix data from \citet{but94} was used in CHIANTI, while at higher temperatures, distorted wave data from \citet[][]{bha06} were used.  }
\label{fig:plot_ecs_081205}
\end{figure}

Both T15 and W17 used the B-spline $R$-matrix method for the scattering calculation. T15, W17, and the present work included 86, 316, and 630 fine-structure target levels, respectively. Fig.~\ref{fig:hexbin_cfecs_081205} shows the hexbin plot comparison of the effective collision strengths at three temperatures in the range of $10^{4-6}$~K. In Fig.~\ref{fig:plot_ecs_081205}, we compare the effective collision strengths of selected transitions listed in Table~\ref{tbl:gml_tbl1}. In general, the data sets agree well. 

%-----------------------------Figure Start----------------------------
%-----------------------------Figure End------------------------------

\subsection{\ion{Ne}{III}}
\label{sct:081003}
The most recent $R$-matrix calculations of electron-impact excitation data for \ion{Ne}{III} (or ${\rm Ne^{2+}}$) are presented in \citet[][M11 hereafter]{mcl11}. Both M11 and the present work used AUTOSTRUCTURE for the atomic structure calculation. As shown in the bottom-right panel of Fig.~\ref{fig:plot_cflev}, the level energies of M11 and the present work agree well with each other. Both are less accurate ($\lesssim10~\%$) compared to NIST. The transition strengths of NIST, M11, and the present work agree well with each other (see the bottom-right panel of Fig.~\ref{fig:plot_cftran}). 

The $R$-matrix ICFT method is used for the scattering calculation of M11 (554 levels) and the present work (630 levels). Fig.~\ref{fig:hexbin_cfecs_081003} shows the hexbin plot comparison of the effective collision strengths at three temperatures in the range of $10^{3-6}$~K. 

\begin{figure}
\centering
\includegraphics[width=\hsize, trim={0.cm 0.5cm 0.5cm 0.5cm}, clip]{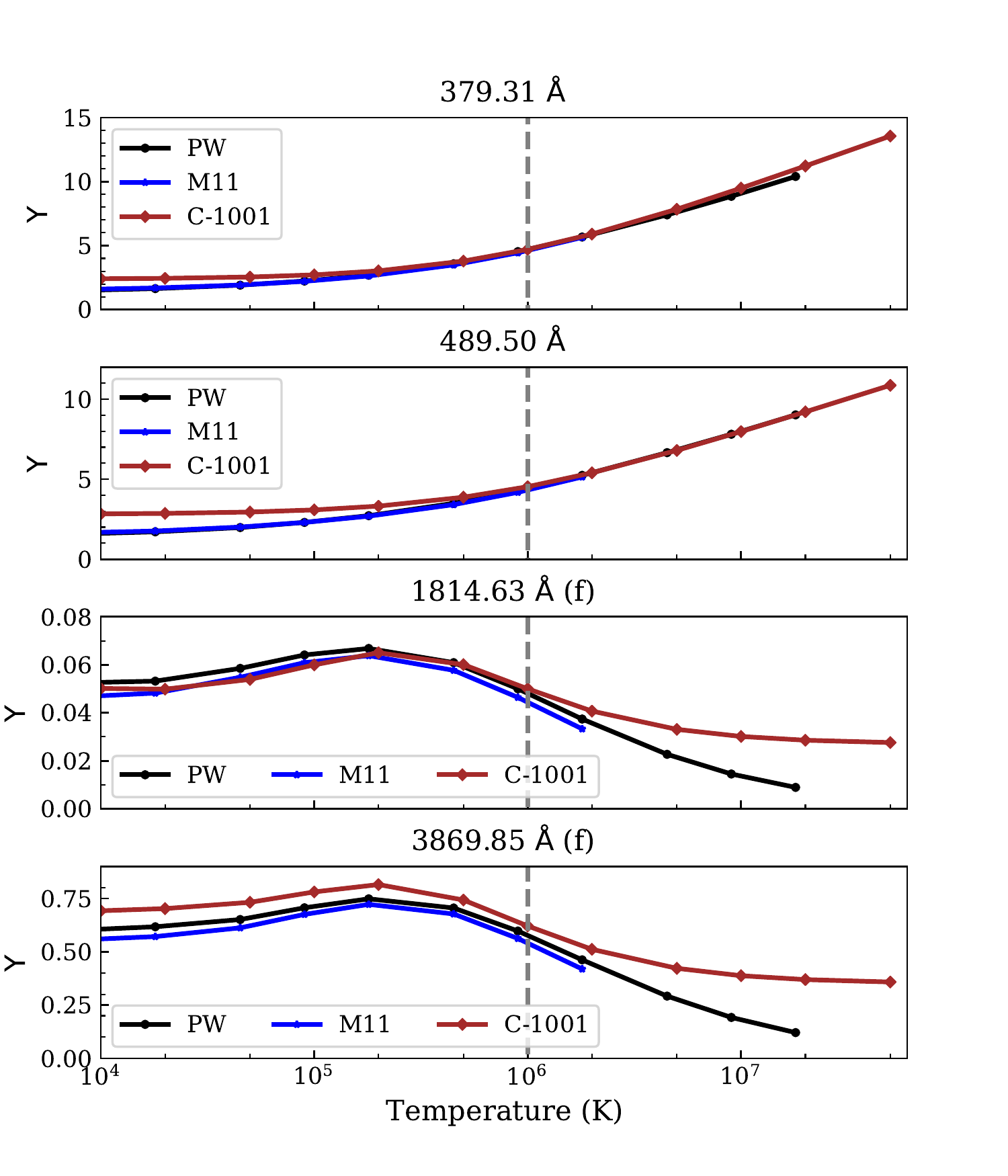}
\caption{Comparison of \ion{Ne}{III} (or ${\rm Ne^{2+}}$) effective collision strengths between the present work (PW), \citet[][M11]{mcl11}, and composite data as incorporated in the CHIANTI atomic database v10.0.1 (C-1001) for selected transitions listed in Table~\ref{tbl:gml_tbl1}. The vertical dashed lines indicate the temperature threshold ($T\sim10^6$~K), below which $R$-matrix data from \citet{mcl00} was used in CHIANTI for transitions within the ground configuration. For other transitions, the distorted wave data from \citet[][]{lan05ne} were used by CHIANTI.}
\label{fig:plot_ecs_081003}
\end{figure}

In Fig.~\ref{fig:plot_ecs_081003}, we compare the effective collision strengths of selected transitions listed in Table~\ref{tbl:gml_tbl1}. Good agreement is found for the allowed transitions, but for the forbidden transitions, the CHIANTI data at $T\gtrsim10^6$~K is systematically larger than the $R$-matrix calculations. 

%-----------------------------Figure Start----------------------------
%-----------------------------Figure End------------------------------

\section{Summary}
\label{sct:sum}
We have presented systematic $R$-matrix intermediate-coupling frame transformation calculations of electron-impact excitation data of O-like ions from \ion{Ne}{III} to \ion{Zn}{XXIII} (i.e. Ne$^{2+}$ to Zn$^{22+}$). For each ion, 630 levels are included in the target configuration interaction and close-coupling collision expansion, which is significantly larger than previous calculations. Level-resolved effective collision strengths are obtained among these levels over a wide temperature range. Since previous $R$-matrix data were available for only some ions, the present work is a significant extension and improvement of electron-impact excitation data of O-like ions. When compared with existing $R$-matrix data in the atomic databases and literature, generally speaking, the new data provided here are consistent within 0.2 dex at temperatures relevant to astrophysical modelling, which is reassuring. When compared to CHIANTI models which used only distorted wave data (e.g., \ion{Ca}{XIII}, \ion{Ar}{XI}, \ion{S}{IX}, and \ion{Si}{VII}), the new data calculated here significantly increase the predicted intensities of many key transitions, and improve or provide new plasma diagnostics which are relevant for current high-resolution spectrometers such as the ground-based DKIST/CryoNIRSP, and the space-based Solar Orbiter/SPICE. 

\begin{acknowledgements}
      The present work is funded by STFC (UK) through the University of Strathclyde UK APAP network grant ST/R000743/1 and the University of Cambridge DAMTP atomic astrophysics group grant  ST/T000481/1. JM thank A. Giunta and R. Dufresne for useful discussion.  
\end{acknowledgements}

%%% References

%\bibliographystyle{aa}
%\bibliography{../bib}  % references}

%-------------------------------------------------------------------

\appendix
\section{Hexbin plot comparisons}
\label{sct:hexbin}
A large number of effective collision strengths are calculated by the present and previous $R$-matrix calculations for some O-like ions. Following \citet{mao20c} and \citet{mao20n}, hexbin plots are used \citep{car87} to compare these results. 

The hexbin plot of Ne {\sc iii} behaves slightly differently than the other ions shown here, exhibiting increased scatter with an increase temperature, from $T=9.00\times10^{4}$~K to $T=1.80\times10^{6}$~K. The scatter is still relatively large (several orders of magnitude) for some transitions. We should point out that the percentage of transitions which differ by at least 0.2 dex ($\sim25$\%) is still comparable to that of Fe {\sc xix} (Table~\ref{tbl:cf_ecs_stat}). If we only focus on transitions from the lowest five energy levels, good agreement is found with the present work (Table~\ref{tbl:cf_ecs_stat}, see also Fig.~\ref{fig:plot_ecs_081003}). 

According to \citet[][M11]{mcl11}, 20 continuum basis orbitals were used by M11 during their calculation to keep the dimensions of the Hamiltonian matrix to a more manageable size. This is because their primary focus was line ratios between transitions within the ground configuration. That is to say, the transitions among high-lying levels are less accurate for M11. 

In the present work, 40 continuum basis orbitals were used for \ion{Ne}{III}. The maximum basis orbital energy (ranging from $14.83$~Ryd for $L=4$ to 20.76~Ryd for $L=19$) covered is a factor of $\sim3.2-4.5$ larger than the ionization potential ($4.66$~Ryd) of \ion{Ne}{III}. We performed an atomic structure calculation using the 24 configurations and the 10 scaling parameters specified in M11. Subsequently, we performed an inner-region exchange calculation with 20 continuum basis orbitals and include angular momenta up to $2J=23$ (i.e. up to $L=19$). In this exercise, the $R$-matrix radius is 22.88 a.u. (as in M11) and the maximum basis orbital energy ranges from $8.04$~Ryd for $L=4$ to 14.72~Ryd for $L=19$, which is a factor of $1.7-3.2$ the ionization potential ($4.66$~Ryd) of \ion{Ne}{III}. 
 
% IP = 4.66 Ryd
% 40 basis orbitals => smallest maximum basis-orbital energy 14.826 Ryd  
% 20 basis orbitals with scaling parameters from M11 for 24 configurations => smallest maximum basis-orbital energy 8.05 Ryd. Accordingly, the maximum basis orbital energy (ranging from $8.04$~Ryd for $L=4$ to 14.72~Ryd for $L=19$, but we don't know whether they stop at $L=19$ or not for the exchange calculation) covered is a factor of $1.7-3.2$ the ionization potential ($4.66$~Ryd) of \ion{Ne}{III}.

%-----------------------------Figure Start----------------------------
\begin{figure*}[b]
\centering
\includegraphics[width=.8\hsize, trim={0.5cm 0.5cm 1.5cm 0.5cm}, clip]{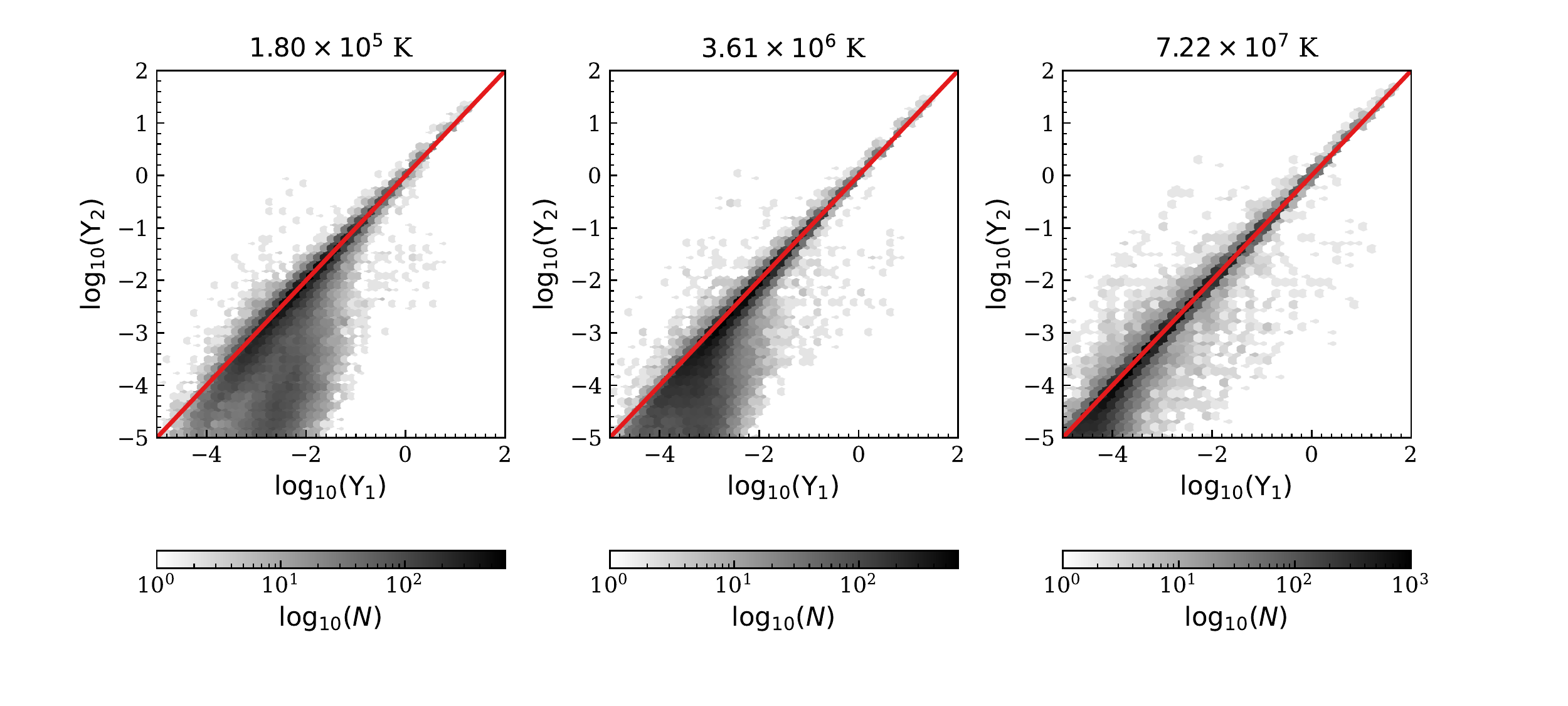}
\caption{Hexbin plots of the comparison of the \ion{Fe}{XIX} (or ${\rm Fe^{18+}}$) effective collision strengths between the present work ($\Upsilon_1$) and \citet[][$\Upsilon_2$]{but08} at $T=1.80\times10^5~{\rm K}$ (left) and $3.61\times10^6~{\rm K}$ (middle), and $7.22\times10^7~{\rm K}$ (right). The darker the colour is, the greater the number of transitions $\log_{10}(N)$. The diagonal line in red indicates $\Upsilon_1=\Upsilon_2$.}
\label{fig:hexbin_cfecs_082619}
\end{figure*}
%-----------------------------Figure End------------------------------

%-----------------------------Figure Start----------------------------
\begin{figure*}
\centering
\includegraphics[width=.8\hsize, trim={0.5cm 0.5cm 1.5cm 0.5cm}, clip]{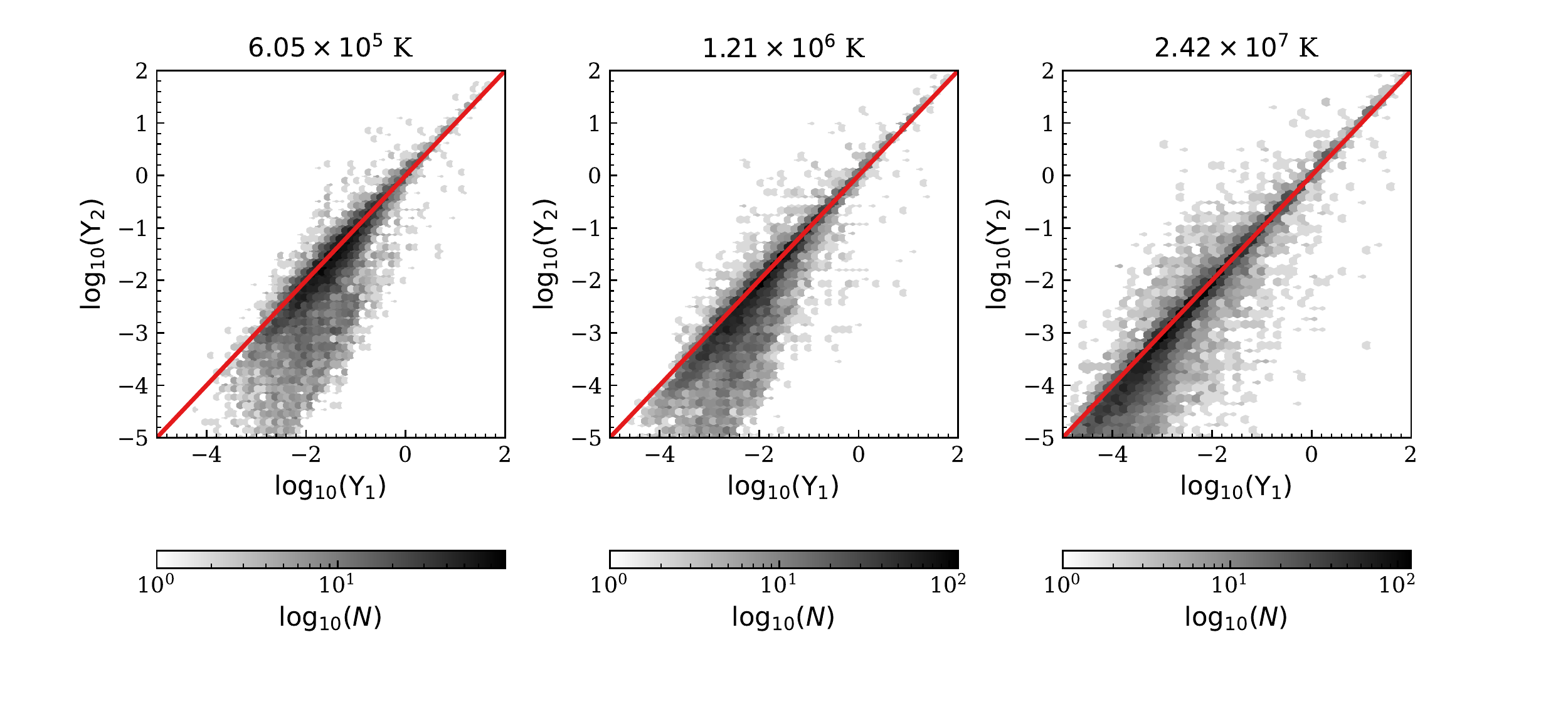}
\caption{Hexbin plots of the comparison of the \ion{Ar}{XI} (or ${\rm Ar^{10+}}$) effective collision strengths between the present work ($\Upsilon_1$) and \citet[][$\Upsilon_2$]{lud10} at $T=6.06\times10^5~{\rm K}$ (left) and $1.21\times10^6~{\rm K}$ (middle), and $2.42\times10^7~{\rm K}$ (right). The darker the colour is, the greater the number of transitions $\log_{10}(N)$. The diagonal line in red indicates $\Upsilon_1=\Upsilon_2$.}
\label{fig:hexbin_cfecs_081811}
\end{figure*}
%-----------------------------Figure End------------------------------

%-----------------------------Figure Start----------------------------
\begin{figure*}
\centering
\includegraphics[width=.8\hsize, trim={0.5cm 0.5cm 1.5cm 0.5cm}, clip]{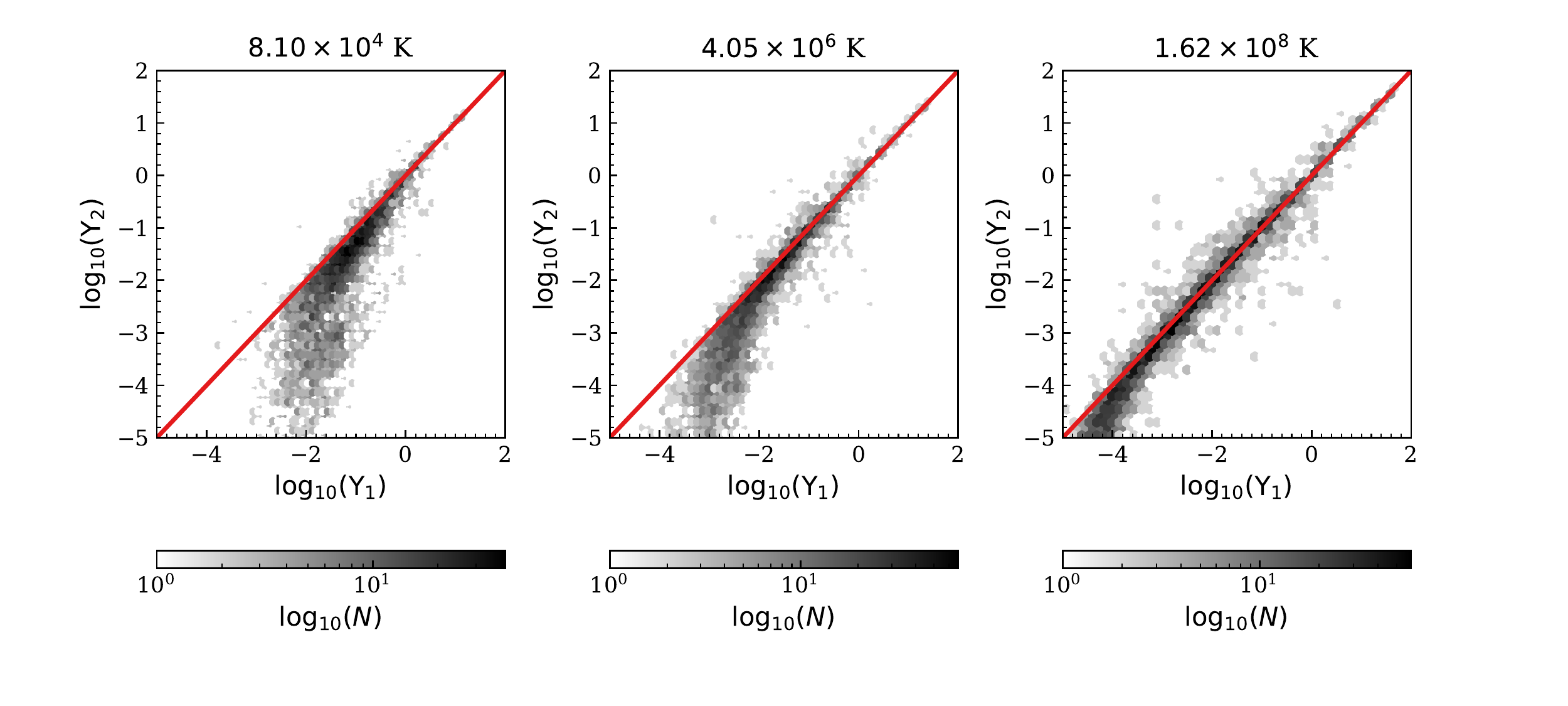}
\caption{Hexbin plots of the comparison of the \ion{S}{IX} (or ${\rm S^{8+}}$) effective collision strengths between the present work ($\Upsilon_1$) and \citet[][$\Upsilon_2$]{lia11} at $T=8.10\times10^4~{\rm K}$ (left) and $4.05\times10^6~{\rm K}$ (middle), and $\sim1.62\times10^8~{\rm K}$ (right). The darker the colour is, the greater the number of transitions $\log_{10}(N)$. The diagonal line in red indicates $\Upsilon_1=\Upsilon_2$.}
\label{fig:hexbin_cfecs_081609}
\end{figure*}
%-----------------------------Figure End------------------------------

%-----------------------------Figure Start----------------------------
\begin{figure*}
\centering
\includegraphics[width=.8\hsize, trim={0.5cm 0.5cm 1.5cm 0.5cm}, clip]{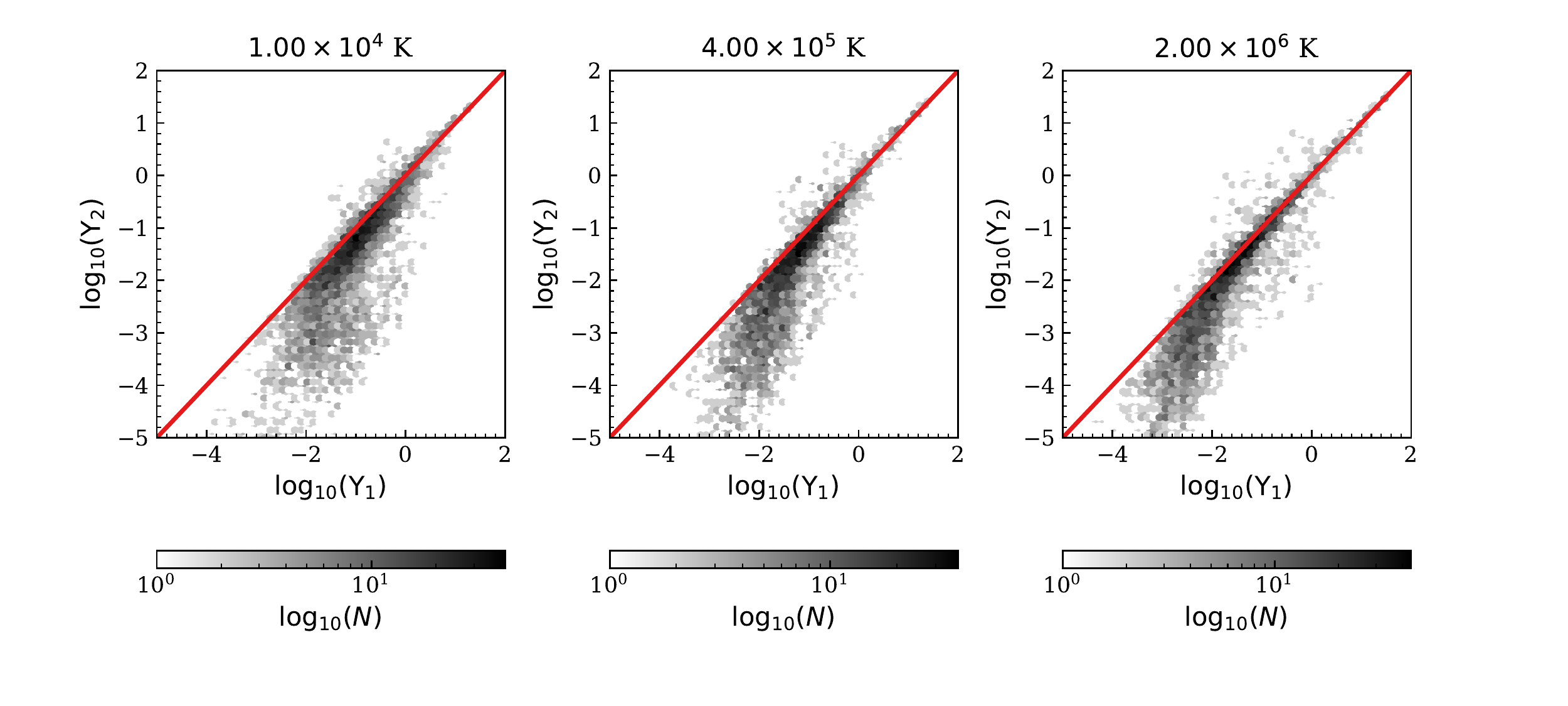}
\caption{Hexbin plots of the comparison of the \ion{Si}{VII} (or ${\rm Si^{6+}}$) effective collision strengths between the present work ($\Upsilon_1$) and \citet[][$\Upsilon_2$]{sos14} at $T=1.00\times10^4~{\rm K}$ (left) and $4.00\times10^5~{\rm K}$ (middle), and $2.00\times10^6~{\rm K}$ (right). The darker the colour is, the greater the number of transitions $\log_{10}(N)$. The diagonal line in red indicates $\Upsilon_1=\Upsilon_2$.}
\label{fig:hexbin_cfecs_081407}
\end{figure*}
%-----------------------------Figure End------------------------------

%-----------------------------Figure Start----------------------------
\begin{figure*}
\centering
\includegraphics[width=.8\hsize, trim={0.5cm 0.5cm 1.5cm 0.5cm}, clip]{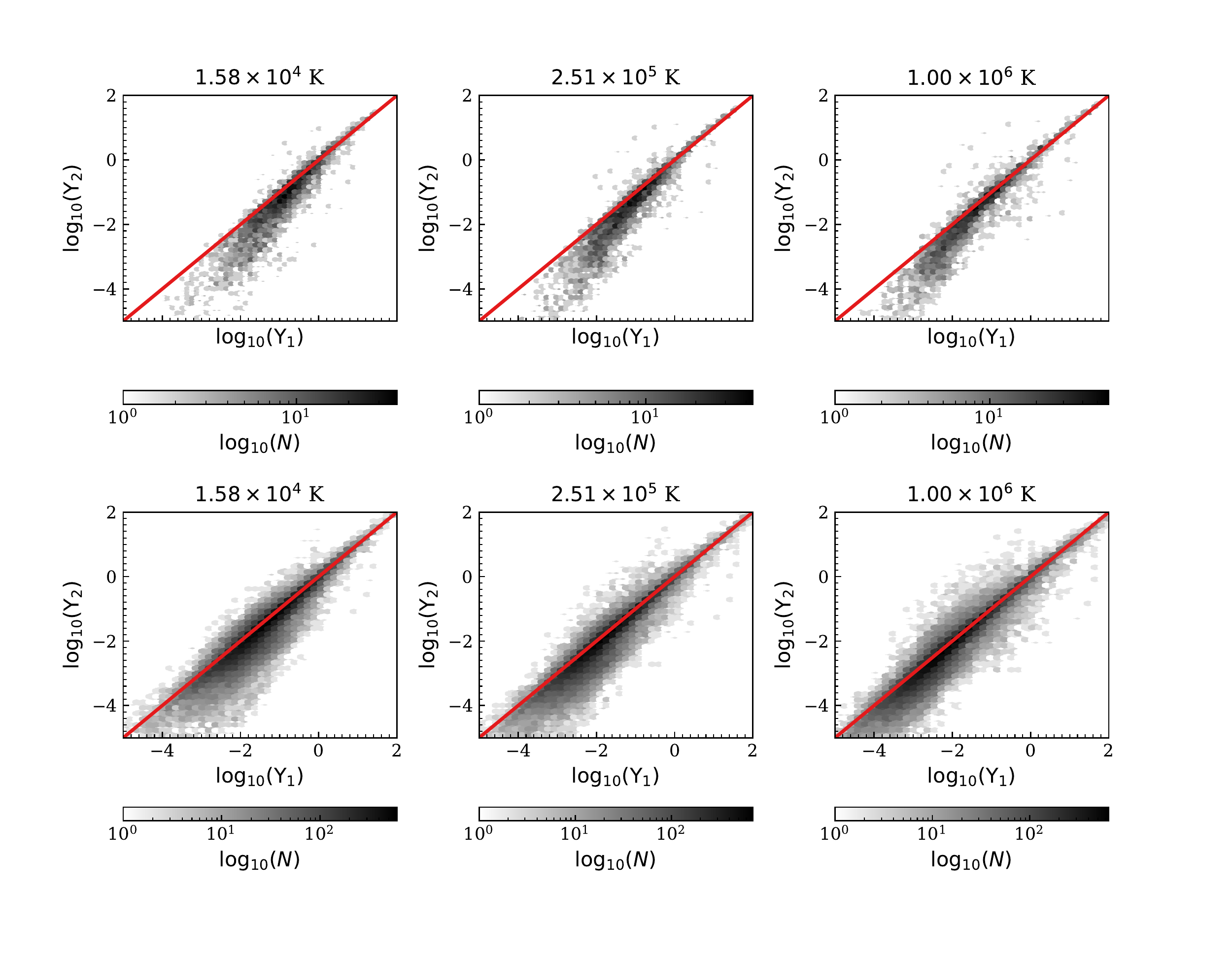}
\caption{Hexbin plots of the comparison of the \ion{Mg}{V} (or ${\rm Mg^{4+}}$) effective collision strengths between the present work ($\Upsilon_1$) and \citet[][$\Upsilon_2$]{tay15} (upper) or \citet[][$\Upsilon_2$]{wan17} (bottom) at $T=1.58\times10^4~{\rm K}$ (left) and $2.51\times10^5~{\rm K}$ (middle), and $1.00\times10^6~{\rm K}$ (right). The darker the colour is, the greater the number of transitions $\log_{10}(N)$. The diagonal line in red indicates $\Upsilon_1=\Upsilon_2$.}
\label{fig:hexbin_cfecs_081205}
\end{figure*}
%-----------------------------Figure End------------------------------

%-----------------------------Figure Start----------------------------
\begin{figure*}
\centering
\includegraphics[width=.8\hsize, trim={0.5cm 0.5cm 1.5cm 0.5cm}, clip]{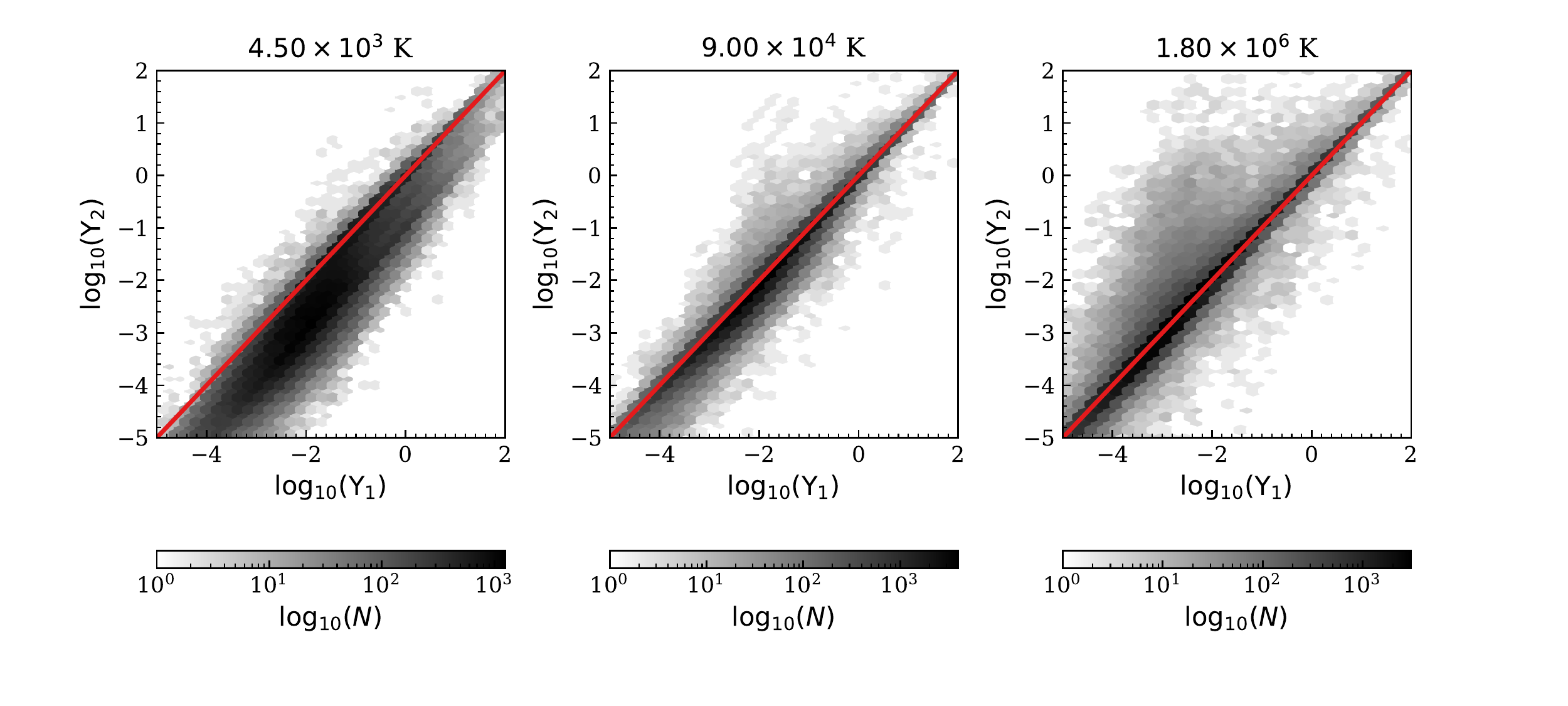}
\caption{Hexbin plots of the comparison of the \ion{Ne}{III} (or ${\rm Ne^{2+}}$) effective collision strengths between the present work ($\Upsilon_1$) and \citet[][$\Upsilon_2$]{mcl11} at $T=4.50\times10^3~{\rm K}$ (left) and $9.00\times10^4~{\rm K}$ (middle), and $1.80\times10^6~{\rm K}$ (right). The darker the colour is, the greater the number of transitions $\log_{10}(N)$. The diagonal line in red indicates $\Upsilon_1=\Upsilon_2$.}
\label{fig:hexbin_cfecs_081003}
\end{figure*}
%-----------------------------Figure End------------------------------

\end{document}